\RequirePackage[2020-02-02]{latexrelease}
\documentclass[prc,aps,preprint,showpacs,nofootinbib,showkeys,eqsecnum,tightenlines]{revtex4}   

\usepackage{graphicx}

\usepackage{dcolumn}

\usepackage{bm}

\usepackage{latexsym,amsmath,amssymb}
\usepackage{amssymb}
\usepackage{amsmath}
\usepackage{amsfonts}
\usepackage{verbatim}

\def\be{\begin{equation}}
\def\ee{\end{equation}}
\def\bea{\begin{eqnarray}}
\def\eea{\end{eqnarray}}
\def\nn{\nonumber}

\newcommand{\noi}{\noindent}

\begin{document}
%
%

\title{Entropy and thermodynamical stability of white dwarfs}
\author{J.~Adam, Jr.\footnote{Corresponding author}\,
\footnote{E-mail address: adam@ujf.cas.cz}}
\affiliation{Institute of Nuclear Physics ASCR, CZ--250 68
\v{R}e\v{z}, Czech Republic }
\author{E.Truhl\'{\i}k\footnote{E-mail address: truhlikemil@gmail.com}}
\affiliation{Institute of Nuclear Physics ASCR, CZ--250 68
    \v{R}e\v{z}, Czech Republic }

\begin{abstract}
A structure of spherical white dwarfs is calculated for a non-zero temperature.
It is shown that the thermodynamic stability of the  white dwarf stars
can be described naturally within the concept of the Helmholtz free energy
of the Coulomb fully ionized electron-ion plasma.
\end{abstract}

\noi 

\noi \hskip 1.9cm 

\maketitle

\section{Introduction}
\label{intro}

Description of the white dwarf stars (WDs) naturally starts from the
equation of equilibrium in the Newtonian approximation, without
rotation and magnetic field included \cite{SC1,MC}\footnote{The
importance of the general relativity for a star with the mass $M$ and
the radius $R$ is given by a compactness parameter $x_g$ \cite{AYP},
\be
x_g\,=\,r_g/R \ , \quad r_g= 2G M/c^2 \approx 2.95 M/M_\odot \ ,
\ee
where $r_g$ is the Schwarzschild radius and $M_\odot$ is the mass of
the Sun. If one takes for $M$ the Chandrasekhar-Landau limit
mass \cite{SC2,LDL} $M\,\approx\,1.4\,M_\odot$ and the radius
$R\,\approx\,5\times 10^3$\,km, one obtains $x_g\,<<\,1$. However,
the effects od the general relativity are important for stabilizing
very massive and fast rotating WDs \cite{SLSSAT}.},
\be
\frac{d\,P}{d\,r}= -\frac{G\,M(r)}{r^2}\,\rho \ , \quad
\frac{d\,M(r)}{d\,r}\,=\,4\,\pi\,r^2\,\rho \ ,    \label{EQLIB1}
\ee
from which it follows,
\be
\frac{1}{r^2}\,\frac{d}{d\,r}
\left(\frac{r^2}{\rho}\,\frac{d\,P}{d\,r}\right)=
-4\,\pi\,G\,\rho \ ,   \label{EQLIB2}
\ee
where $P$ is the pressure, $\rho$ is the mass density and $G$ is the Newton
gravitational constant. One gets explicitly from  (\ref{EQLIB2}),
\be
\frac{1}{\rho}\,\frac{d^2\,P}{d\,r^2}+ \frac{2}{\rho\,r}\,
\frac{d\,P}{d\,r}\,-\,\frac{1}{\rho^2}\,\frac{d\,P}{d\,r}\,\frac{d\,\rho}{d\,r}
+ 4\,\pi\,G\,\rho= 0 \ .   \label{EQLIB3}
\ee
If  $P$ as a function of $\rho$  (i.e., an equation of state (EoS)) is known, the equation
above can be cast into a second order differential equation for
the function $\rho(r)$.

For instance, it was shown in \cite{SC1} that assuming the EoS in
the form
\be
P= K\,\rho^{(1+\frac{1}{n})} \ ,   \label{EOS1}
\ee
and writing $\rho$ as
\be 
\rho= \rho_c\,\theta^n \ ,      \label{RHON} 
\ee
one gets EoS eq.\,(\ref{EOS1}) parameterized as,
\be P= K\, \rho_c^{(1+\frac{1}{n})}\,\theta^{(1+n)} \ .
\label{EOS2} \ee
Then, from eq.\,(\ref{EQLIB3}) the famous Lane-Emden (LE) equation follows,
\be
\frac{1}{\xi^2}\,\frac{d}{d\,\xi}\,\left(\xi^2\,\frac{d\,\theta}{d\,\xi}\right)
= 0 \ ,   \label{LEEQ}
\ee
where 
\be 
\xi= \frac{r}{a} \ , \quad a= \sqrt{\frac{(n+1)}{4\,\pi}\,
\frac{K}{G}\, \rho_c^{\left(\frac{1}{n}-1\right)} } \ .
\label{XIA} 
\ee
Further, usually one sets $\rho_c= \rho(0)$, i.e.  $\rho_c$ is the central mass density. Then the initial conditions are $\theta(0)= 1$ and
$\left(d\,\theta/d\,\xi\, \right)|_{\xi=0}= 0$.\\

Equation~((\ref{LEEQ}) was discussed in detail in \cite{SC1}, where
it was shown that for $n= 0,\, 1,\, 5$ one gets explicit solutions.
The model with the EoS (\ref{EOS2}), known as polytropic model, has
been frequently used in the astrophysics, in particular for treating
the structure of WDs. In this case, the star is considered as a
dense Coulomb plasma of nuclei in a charge compensating background
of degenerate electron Fermi gas, where the electron-ion interaction
is taken in this polytropic approximation. However, for studying the
WDs structure, the polytropic model is of restricted use, because it
fairly approximates the equation of state only in the extreme
\mbox{non-relativistic} limit for $\rho_c << 10^6\, $g/cm$^3$ (with
$n$= 3/2)  and in the extreme relativistic limit for $\rho_c >>
10^6\, $g/cm$^3$ (with $n$= 3).
Recall that  for $n= 3$ the mass of the WDs is uniquely given by
the Chandrasekhar-Landau limit mass \cite{SC2,LDL}.

Another problem with the polytropic model is its stability. As
discussed at length in \cite{JPM}, the squared
Brunt-V\"{a}is\"{a}l\"{a} frequency $N^2\,=\,0$ in this model. Only
if $N^2\,>\,0$, the fluid is stable, if $N^2\,<\,0$, it is
convectively unstable. So the fluid described by the polytropic
model is only neutrally stable. According to Ref.~\cite{JPM}, no
magnetic equilibrium exists in this case.

Stable stratification has an important influence on stellar magnetic
equilibria and their stability.  As discussed at length in
Ref.~\cite{TARMM}, the study of the magnetic equilibria in stable
stratified stars has a long story. It follows from these studies
that simple magnetic field configurations are always unstable. In
their study, the authors of  Ref.~\cite{TARMM} constructed simple
analytic models for axially symmetric magnetic fields, compatible
with hydromagnetic equilibria in stably stratified stars, with both
poloidal and toroidal components of adjustable strength, as well as
the associated pressure, density and gravitational potential
perturbations, which maked them possible to study directly their
stability. It turned out  that the instability of toroidal field can
be stabilized by a poloidal field containing much less energy than
the former, as given by the condition  $E_{pol}/E_{tor} \gtrsim 2 a
E_{pol}/E_{grav}$,  where $E_{pol}$ and
$E_{tor}$  are the energies of the poloidal and toroidal
fields, respectively and $E_{grav}$ is the  gravitational
binding energy of the star. It was found in \cite{TARMM} that $a \approx 7.4$
for main-sequence stars, which compares with $a \approx 10$  obtained by Braithwaite \cite{JB},
using the method of numerical simulations.  But the  results for the neutron stars differ
by a factor of $\approx$ 4. The possibility of compensation of instabilities of toroidal fields
by a relatively weak poloidal field was earlier studied by Spruit \cite{HCS}.
As to the instabilities in the poloidal field, stable stratification
is of less help for eliminating them \cite{TARMM}. A relatively
stronger toroidal field would be needed in order to stabilize it \cite{JB}.


As it follows from the discussion in Ch.~3 of the monograph
\cite{CSBK}, the necessary and sufficient condition for the
thermodynamical stability of stars with optically thick media is the
positive gradient of the entropy,
\be
\frac{d\,S}{d\,r} > 0 \ .       \label{PGE}
\ee
Respecting this criterion of the stability, the authors of Ref.~\cite{JPM} considered the star as
a chemically uniform, monoatomic, classical ideal gas, described in the polytropic model
with the EoS (\ref{EOS1}) as $P\,\sim\,\rho^\gamma$, where
$\gamma= 4/3=  1+ 1/n \ , (n= 3)$ and with the specific
entropy
$$
s\approx ln(P/\rho^\Gamma)+ const \ , \qquad {\rm where} \quad
\Gamma= \left(\frac{\partial\,P}{\partial\,ln\,\rho}\right)_{ad}= \frac{5}{3} \ ,
$$
for which it holds
\be
\frac{d\,s}{d\,r}  > 0 \ .    \label{PGE1}
\ee
In this model, applied to the Ap stars, the constructed  magnetic
equilibrium turns out to be stable\footnote{In our opinion,  it would
be more consistent for Mitchell et al. \cite{JPM}  to consider the
EoS of the classical ideal gas,  $P\,\sim\,\rho^\gamma\,T$, with
$\gamma= 1$, for which Eq.~(\ref{PGE1}) also holds.}.  However,
for $\gamma= 5/3 \ , (n= 3/2)$, one obtains
$\gamma= \Gamma \ , \ d\,s/d\,r= 0$ and the magnetic equilibrium
is unstable. Similar calculations have recently been done in Ref.\,\cite{LB}.

This model cannot be applied to the study of WDs, because they
consist of plasma containing the mix of the fully ionized atoms and
of the fully degenerate electron gas.

The polytropic model was used to describe super-Chandrasekhar
strongly magnetized  WDs (SMWDs) in Refs.~\cite{UDM1,UDM2,UDM3} and also in Ref.~\cite{BB}.
It was shown in \cite{BB}, that axisymmetric configurations with the poloidal or
toroidal fields are unstable and it was concluded that the long
lived super-Chandrasekhar SMWDs are unlikely to occur in nature.

In Ref.~\cite{DC}, the authors developed a detailed and
self-consistent numerical model with a poloidal magnetic field
in the SMWDs. In their model, the rotation, the effects of general
relativity and a realistic EoS were taken into account and extensive
stability analysis of such objects was performed. As a result, it
was found that the SMWDs could potentially exist, but their
stability would be mainly limited by the onset of the electron
capture and of the picnonuclear reactions. However, it should be
noted that the condition of the thermodynamical stability
(\ref{PGE}) was not checked in Ref.~\cite{DC}.

In the recent paper \cite{NCLP}, the authors have studied the
influence of the electron capture on the stability of the WDs. They
used the EoS in the polytropic form (\ref{EOS1}) in the
ultrarelativistic limit $P\approx K_0\,\rho^{4/3}$, with $A$ and
$Z$ dependent constant $K_0$. Besides, the electrostatic correction
and the correction due to the electron-ion interaction are  also
included in $K_0$. This allowed them to calculate the threshold
density $\rho_\beta$ for the capture of electrons and to set the
lower bound for the radius of the WDs. It was also found that the
electron capture reduces the mass of WDs by 3 - 13\%. Solving the
Einstein-Maxwell equations, with the magnetism of the dense matter
included, has shown that the magnetized WDs with the polar magnetic
field stronger than $10^{13}\, $G could be massive enough to explain
overluminous type Ia supernova. It was also found that the pure
poloidal  magnetic field is unstable. Actually, this result follows
from the fact that the polytropic model is only neutrally stable, as
it has already been discussed above.

In Ref.~\cite{LB1}, the authors investigated the evolution of
isolated massive, highly magnetized and uniformly rotating WDs,
under angular momentum loss driven by magnetic dipole braking. They
computed general relativistic configurations of such objects using
the relativistic Feynman-Metropolis-Teller equation of state
for the WD matter. One of the interesting results is obtained for
rotating magnetized WD with the mass which is not very close to the
non-rotating Chandrasekhar-Landau mass. Such a WD evolves
slowing down and behaving as an active pulsar of the type of soft
gamma-repeater and anomalous X-ray pulsar \cite{MM,IB,KBLI,JAR,JGC,RVL,VBB,TRM}.
Let us note that it is not clear if the condition of
the thermodynamical stability (\ref{PGE}) is fulfilled in
Ref.~\cite{LB1}.

In Ref.~\cite{Dex} the effect of magnetic field in hot WDs is studied,  modeling
their interior by a lattice of nuclei surrendered by a relativistic free Fermion electron gas.
This study is based  on the electron pressure as a tensor, an approach which was criticized in Ref.~\cite{PY}.   The response to this criticism has been published in 
Ref.~\cite{Dex2}. Still, it is not clear, if the electron entropy in eq.~(5) of Ref.~\cite{Dex}
satisfies the stability condition  $ds/dr > 0$.

Realistic model of magnetized fully-ionized
plasma has been developed in \cite{CP,BPY,PC1,PC2,PC3,ASJ}. In the model,
considered in \cite{CP,BPY,PC1,PC2,PC3,ASJ}, an analytical EoS is derived
from the Helmholtz free energy of the system of magnetized
fully-ionized atoms and of the degenerate electron gas, whereas,
in \cite{ASJ} also positrons were included. Such an EoS covers a
wide range of temperatures and densities, from low-density classical
plasmas to relativistic, quantum plasma conditions.

Starting in Sect.~\ref{eqstab} from the equation of equilibrium
(\ref{EQLIB3}), and using in Sect.~\ref{hfeaeos} the function
$P(\rho)$, obtained from the EoS mentioned above, we get the second
order equation for the matter density $\rho$. Solving this equation,
we study the structure of corresponding WDs for a representative
series of values of the central density $\rho_c$, chosen from the interval
$10^4\,\mathrm{g/cm^3} << \rho_c << 2\times\,10^9\,\mathrm{g/cm^3}$,
and simultaneously using the electron and ion entropies from
Sects.~\ref{eentropy} and \ref{ientropy}, we
show that the criterion of the thermodynamic
stability (\ref{PGE})  is fulfilled in all cases (see Fig.~2 and Fig.~3).
Besides, using LE eq.~(\ref{LEEQ}), we calculate for comparison for the
extreme non-relativistic and extreme relativistic  values of
$\rho_c$ the mass and radius of corresponding WDs, which are
presented in Table~\ref{Tab1}.

In Appendix A, we discuss different ways of calculating
the structure of WDs, specify numerical values of our input parameters,  and summarize relations for the scaling parameter  $a_s$ and for the LE approximation. In Appendix B, we express  the functions $f_1$ and $f_2$, entering the pressure $P(\rho)$, in terms of the Fermi-Dirac integrals
and in Appendix C,  we briefly describe how to decompose the thermodynamic
quantities for free electrons into series in powers of $k_B T/\tilde{E}_F$, where
$\tilde{E}_F= \mu_e(T=0)$  is the Fermi energy with the rest mass contribution subtracted.

Our results show that the realistic model developed in
\cite{CP,BPY,PC1,PC2,PC3,ASJ}
is a good starting model to be applied to construct the WDs with stable magnetic fields.

\section{Methods and input}

\subsection{Modified equation of stability}
\label{eqstab}

Let us write eq.~(\ref{EQLIB3}) in the form,
\be
\frac{d^2\,P}{d\,r^2}+ \frac{2}{r}\, \frac{d\,P}{d\,r}-
\frac{1}{\rho}\,\frac{d\,P}{d\,r}\,\frac{d\,\rho}{d\,r}
+ 4\,\pi\,G \,\rho^2= 0 \ .   \label{EQLIB4}
\ee
Considering now the pressure $P$ as a function (solely) of the
density $\rho$, eq.\,(\ref{EQLIB4}) can be transformed into the
2nd order differential equation for $\rho$,
\be
\frac{d^2\,\rho}{d\,r^2}+ f_1(\rho)\,\left(\frac{d\,\rho}{d\,r}\right)^2
+ \frac{2}{r}\,\frac{d\,\rho}{d\,r}+ \frac{4\,\pi\,G}{f_2(\rho)}\,\rho^2=
 0 \ ,  \label{EQFRHO}
\ee
where
\begin{equation}
f_1(\rho)= \left(\frac{d^2\,P}{d\,\rho^2}\right)\bigg/
\left(\frac{d\,P}{d\,r}\right)- \frac{1}{\rho} \ , \qquad
f_2(\rho)= \frac{d\,P}{d\,\rho} \ .               \label{f1f2a}
\end{equation}
Next we set
\be 
r=a_s\,x \ , \quad \rho= \rho_c\, y \ .     \label{xy}
\ee
Since $\rho_c$ is the matter density in the center of the star,
\be 
y(0)= 1 \ , \quad \frac{d\,y}{d\,r}\big|_0= 0 \ .   \label{bc}
\ee
From relations
\be 
\frac{dP}{d\rho}= \frac{dP}{\rho_c \,d y} \ , \quad
\frac{d^2P}{d\rho^2}\, = \frac{d^2P}{\rho_c^2 \,d y^2} \ ,
\label{dpdy} 
\ee
it follows
\be 
f_1(\rho)= \frac{1}{\rho_c}\,f_1(y) \ , \quad f_2(\rho)=
\frac{1}{\rho_c}\,f_2(y) \  .    \label{f1f2b} 
\ee
Then, in terms of the new variables (\ref{xy}), eq.\,(\ref{EQFRHO}) becomes,
\be
\frac{d^2\,y}{d\,x^2}+
f_1(y)\,\left(\frac{d\,y}{d\,x}\right)^2+
\frac{2}{x}\,\frac{d\,y}{d\,x}+ \frac{C}{f_2(y)}\,y^2= 0 \ ,  \label{EQFY}
\ee
with
\be 
C= 4\,\pi\,G\,a_s^2\,\rho_c^2 \ .     \label{C} 
\ee
We solved eq.\,(\ref{EQFY}) by the standard 4th order Runge-Kutta
method for various values of the central matter density $\rho_c$.
The choice of the value of the scaling parameter $a_s$ is discussed
in Appendix A. In our calculations we use the value (see
(\ref{myRs}))
\be
a_s= 8686.26\, {\rm km} . \label{scale}
\ee

\subsection{The Helmholtz free energy and the EoS}
\label{hfeaeos}

Here we follow Ref.~\cite{PC3}, where the  Helmholtz free energy
$F$ of the plasma is defined as,
\bea
F &=& F^{(e)}_{id}+ F^{(i)}_{id}+ F_{ii}+ F_{ee}+ F_{ie} \ , \nn\\
&\simeq& F^{(e)}_{id}+ F_{lat}  \ , \quad
F_{lat}= F^{(i)}_{id}+ F_{ii} \ . \label{HFE}
\eea
On the first line, the first two terms correspond to the ideal part
of the free energy of ions and electrons, and the last three terms
correspond to the ion-ion, electron-electron, and
ion-electron interactions, respectively. The second line
corresponds  to the approximation adopted in this paper: the sum of
the 2nd and 3rd terms of the 1st line is denoted as $F_{lat}$ and
evaluated as in Ref.\,\cite{PC3} in one-component plasma model
(OCP) in the regime of Coulomb crystal. Less important and more
uncertain  contributions $F_{ee}$ and $F_{ie}$ are skipped. It
should be noted that these contributions are less important only
in degenerate plasmas.  In the non-degenerate regime, they can be of the
same order of magnitude as $F_{ii}$, especially if $Z$ is not large \cite{PC3}.
\begin{figure}
        \includegraphics[width=11.0truecm]{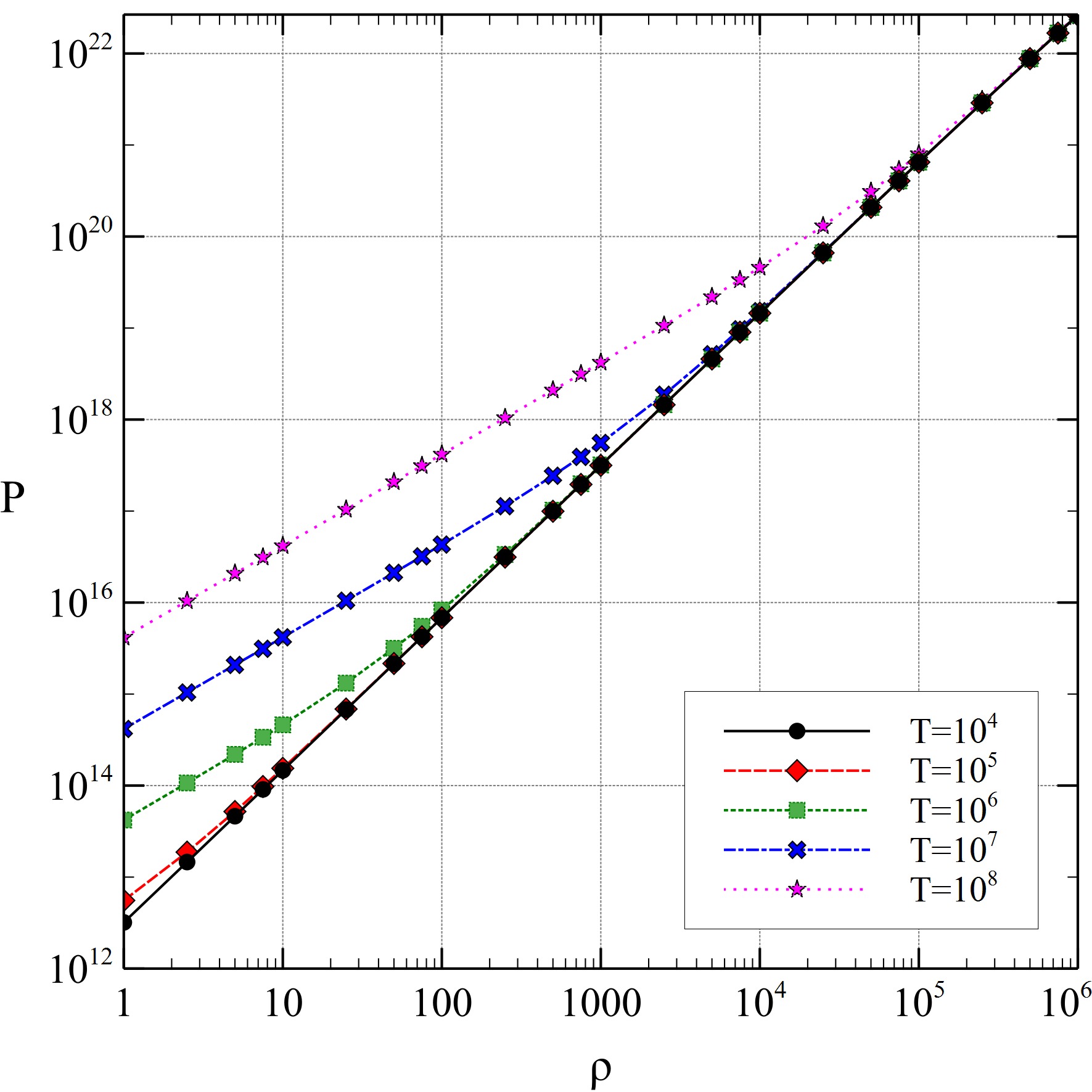}
        \includegraphics[width=11.0truecm]{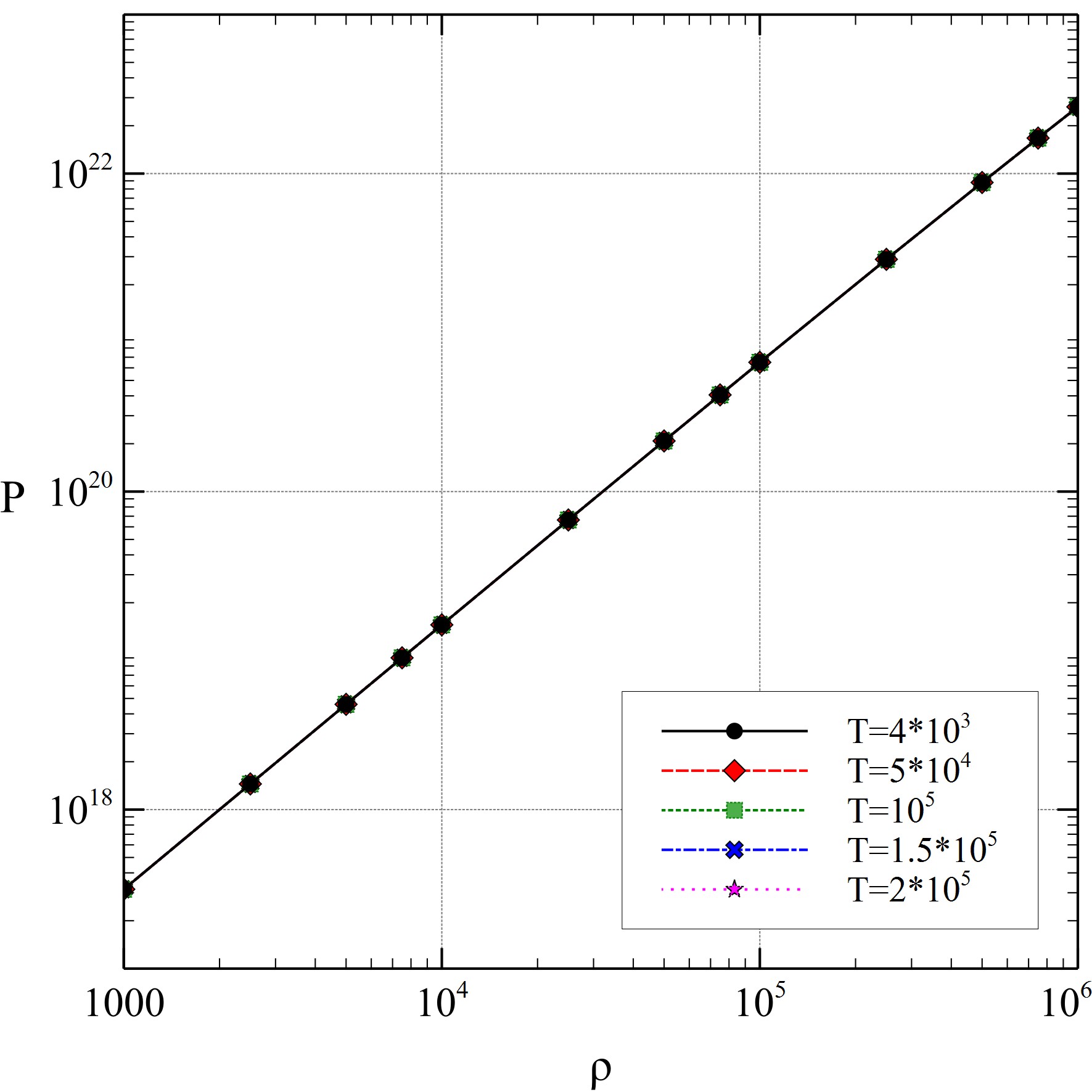}
\caption{Dependence of the electron pressure
$P^{(e)}\, $ [g/cm$\cdot {\rm sec^2}]$
on the matter density $\rho\, $[g/${\rm cm^3}]$ and the temperature $T\, $[K].
In the upper panel, somewhat wider ranges of densities and temperatures
are considered, while in the lower one, they are relevant for the carbon WDs.}\label{fig:rhoP}
\end{figure}

The particle density $N$,  internal energy $U$, the entropy $S$
and Helmholtz free energy $F$ are related by:
\bea
T\, S &=& U+ P\, V- \mu\, N \ , \label{defS}\\
F &=& U- T\, S=  \mu\, N -  P\, V  \ . \label{defF}
\eea

In the lower panel of Fig.~1, we present the dependence of the electron pressure on
$\rho$ for various values of $T$ for the carbon WDs with $A=12,\, Z= 6$.
In this case, the dependence of the electron pressure on the matter density is $T$-independent
\footnote{Similar $ P^{(e)}-\rho$ dependence on
    various values of $T$ was presented in Fig.~1 of Ref.~\cite{KB}.}.

In general,  the pressure and the entropy can be calculated from $F$ by:
$$
P= -\left( \frac{\partial\,F}{\partial\,V} \right)_T \, ,  \qquad
S= -\left( \frac{\partial\,F}{\partial\,T} \right)_V \ .
$$
But for free electrons the number density $n_e$, pressure
$P^{(e)}_{id}$  and energy $ U^{(e)}_{id}$ are known explicitly:
\bea
 \label{ne}
n_e &\equiv& \frac{N_e}{V}=
c_n\,\big[I_{1/2}(\chi_e, \tau)+ \tau\, I_{3/2}(\chi_e , \tau)\big]\ , \\
 \label{peid}
  P^{(e)}_{id} &=&
c_p\,\big[I_{3/2}(\chi_e, \tau)+ \frac{\tau}{2}\, I_{5/2}(\chi_e , \tau)\big]\  , \\
 \label{Ueid}
  U^{(e)}_{id} &=& c_e\,\big[I_{3/2}(\chi_e, \tau)+
\tau\, I_{5/2}(\chi_e, \tau)\, \big]\equiv {\cal E}_e\, V \ ,
\eea
where $\chi_e= \mu_e\, \beta$, $\mu_e$ is the electron
chemical potential without the rest energy $m_ec^2$ and dimensionless
$\tau= T/T_r\, ,$ with $T_r= m_e\,c^2/k_B\simeq
5.9301 \times 10^9\, $K (from the Boltzmann constant
$k_B \simeq 8.617\times 10^{-11}\, $MeV/K). In the last relation
we introduce the electron energy density ${\cal E}_e=  U^{(e)}_{id}/V$.
The electron energy $ U^{(e)}_{id}$ and its density ${\cal E}_e$ also do
not contain the rest mass contributions.
Further, using the electron Compton length $\lambda_e$  from (\ref{Lame}):
\bea
c_n &=&
\frac{\sqrt{2}}{\pi^2\hbar^3}\left(\frac{m_e}{\beta}\right)^{3/2}
= 3\sqrt{2}\rho_0\, \tau^{3/2} \ , \quad \beta= \frac{1}{k_B T} \ , \quad
\rho_0= \frac{1}{3\pi^2\, \lambda_e^3} \ , \nn\\
c_p &=& \frac{(2 m_e)^{3/2}}{3\pi^2\hbar^3\beta^{5/2}}=
2\sqrt{2}\, m_ec^2\, \rho_0\, \tau^{5/2} \ , \nn\\
c_e= \frac{3}{2}c_p &=& \frac{\sqrt{2}\, m_e^{3/2}}{\pi^2\hbar^3\beta^{5/2}}=
3\sqrt{2}\, m_ec^2\, \rho_0\, \tau^{5/2} \ . \nn
\eea

The generalized (relativistic) Fermi-Dirac integrals
$I_{\nu}(\chi_e,\,\tau)$ are defined as follows:
\be
I_{\nu}(\chi_e\, , \tau) =
\int\limits_0^\infty\, \frac{x^\nu
\, \sqrt{1+ \tau\, x/2\, }\, }
{\mathrm{e}^{(x- \chi_e)}\,+\,1}\, dx\ . \label{FDI}
\ee
The functions $f_1(y)$ and $f_2(y)$ of previous section are
presented in terms of the Fermi-Dirac integrals
$I_{\nu}(\chi_e,\,\tau)$ in Appendix B.\\

One obtains the EoS by referring to the neutrality of the plasma,
which provides the equation between the mass density $\rho$ of the
ion and the electron number density $n_e$,
\be
n_e= Z n_i= \frac{Z \rho}{A m_u}= \frac{\rho}{\mu_u} \ , \label{neut}
\ee
where $Z$ is the ion charge number, $A$ is the mass number,
$n_i$ is the ion number density, \\ $m_u$ is
the atomic mass unit and $\mu_u= A\, m_u/Z$ is introduced in
(\ref{muu}). Given values of $A,\, Z$ and $\rho$ one gets $n_e$ from
the neutrality condition (\ref{neut}) and then reversing eq.~(\ref{ne})
for a given temperature $T$, obtains the value of $\chi_e$.
Substituting this $\chi_e$
into eq.~(\ref{peid}), one gets the value of the electron pressure
$ P^{(e)}_{id}$ corresponding to the given value of $\rho$.

\begin{table}[b]

\caption{\label{Tab1} In the upper part  the results of calculations
of the radii and masses of the WDs for representative values of the
central mass densities $\rho_c$ are presented. The equation of
stability~(\ref{EQFY}) was solved with the scaling parameter $a_s$
of eq.~(\ref{scale}). The EoS is obtained from the Helmholtz free
energy of the Coulomb plasma, using eqs.~(\ref{peid}),\,(\ref{ne})
and (\ref{neut}). The central fractional Fermi momentum
eq.~(\ref{xr}) $x_{Fc}$ is also listed, since it indicates whether
the dynamics is non-relativistic or relativistic. In the lower part
of the table (below the middle double line) we present the radii and
masses obtained from  the LE eq.~(\ref{LEEQ}), which is based on the
polytropic EoS eq.~(\ref{EOS2}). $R_0$ and $M_0$ are radii and
masses in this approximation, $r_0$ and $m_0$ the corresponding
dimensionless quantities, introduced at the end of Appendix~A in
eqs.~(\ref{r0nr}-\ref{mmsunnr}) and (\ref{r0rel}-\ref{mmsunrel}).
Comparing these (LE) results with our numerical ones (which do not
employ a polytropic EoS) one can see that they are close to each
other in the non-relativistic case, but in the relativistic regime
they approach each other only in the extreme relativistic limit of
very large densities.} \label{tab-radmas}
\begin{ruledtabular}
\begin{tabular}{|c||c|c|c|c|c|c|}
model & 1 & 2 & 3 & 4 & 5 & 6   \\  \hline
$\rho_c {\rm [g/cm^3]}$ & $10^4$ &  $10^5$ & $ 10^6 $ & $10^7$ &
$10^8$ & $2\cdot 10^9$ \\ \hline
 $x_{Fc}$ & $ 0.173 $ &  $ 0.372 $ & $ 0.801 $ &
 $ 1.36  $ & $ 3.72 $ & $ 10.1 $ \\ \hline
$R_{WD}$ [km] & $2.40\cdot10^4$ &
$1.63\cdot10^4$ & $1.09\cdot10^4$ & $7.04\cdot10^3$ &
$4.30\cdot10^3$ & $2.05\cdot10^3$  \\ \hline
$M/M_\odot$  & 0.048 &  0.146 & 0.391 & 0.816 & 1.15 &
1.37 \\ \hline \hline
 n  & $3/2$ &  $3/2$ & & 3 & 3 & 3 \\ \hline
  $r_0$ & $ 2.78 $ &  $ 1.90 $ & $  $ &
 $ 1.79  $ & $ 0.830 $ & $ 0.306 $ \\ \hline
 $m_0$ & $ 0.0185 $ & $ 0.0583 $ & $  $ &
 $ 0.542  $ & $ 0.542 $ & $ 0.542 $ \\ \hline
$ R_0 $ [km] & $2.42\cdot 10^4$ &
$1.65\cdot 10^4 $ & &
 $1.55\cdot 10^4 $ & $7.21\cdot 10^3 $ & $2.66\cdot 10^3$ \\ \hline
$M_0/M_\odot$  &  0.050 & 0.157 & & 1.46 & 1.46 & 1.46 \\
\end{tabular}
\end{ruledtabular}
\end{table}


Apart from  the electron pressure, Chamel and Fantina \cite{CHFA}
have taken into account also the lattice pressure $P_{lat}$,
derived from the dominant static-lattice (Madelung) part of the ion
free energy \cite{PC3}, i.e. approximating $F_{lat}\simeq F_M$:
\be
F_M = N_i\, k_B\, T\, C_M\, \Gamma \ ,
\quad N_i= n_i\, V \ , \label{FM}
\ee
and for the bcc crystal  the Madelung constant is \cite{BPY}:
\bea
 C_M &=& -0.89592925568 \ . \label{cmad}
\eea
The ion coupling parameter
\be
\Gamma= \frac{(Z e)^2}{a_i k_B T} \ , \label{Gioncoup}
\ee
is given in terms of the ion sphere radius
$a_i= \big(\frac{4}{3} \pi n_i\big)^{-1/3}$.

As it can be seen from Table III of Ref.~\cite{CHFA}, the effect of the
pressure $P_{lat}$ on the mass of the WDs containing the
light elements is only few percents and we do not take it into account.
The results of comparative calculations are presented in Table~\ref{Tab1}.

\section{Entropy and its gradient in WDs}
\label{entropy}

In this section we calculate one-electron and ion entropies and
their gradients within the above introduced Coulomb plasma theory
based on the Helmholtz free energy concept. We show on the
representative set of WDs that both entropies are positive and that
their gradients satisfy the condition of the thermodynamical
stability, required by eq.~(\ref{PGE}). We will deal with a reduced
dimensional entropy defined (both for electrons and ions) as:
\bea
\hat{s} &=& \frac{S}{k_B\, N} \ , \label{reducS}
\eea
where $N$ is a number of particles and  $k_B$ is the
Boltzman constant. We will evaluate and plot a derivative of various contributions to
this reduced entropy in respect to the dimensionless radius $x$ (see (\ref{xy})), i.e.,
$d\hat{s}/d\, x$.

\subsection{The electron entropy}
\label{eentropy}

For the free electrons it follows from (\ref{reducS}) and relations
(\ref{ne}-\ref{Ueid}):
\bea
\hat{s}_e &=& \frac{1}{k_B T}
\frac{U^{(e)}_{id}+ P^{(e)}_{id}\, V-
N_e\, \mu_e}{V\, n_e}=
 \frac{1}{k_B T}\, \left( \frac{{\cal E}_e+
    P^{(e)}_{id}}{n_e}- \mu_e \right) \nn\\
&=& \frac{5\, I_{3/2}(\chi_e,\tau)+ 4\tau\, I_{5/2}(\chi_e,\tau)}
  {3(I_{1/2}(\chi_e,\tau)+ \tau\, I_{3/2}(\chi_e,\tau))}
 - \chi_e \simeq
 \pi^2\, \tau\, \frac{\epsilon_F}{x^2_F} \ ,   \label{hats}
\eea
where $n_e$ is the electron density,
 $\tau= T/T_r, \, T_r= m_ec^2/k_B$. Further, $p_F= \hbar (3 \pi^2 n_e)^{1/3}$
is  the electron Fermi momentum and the dimensionless Fermi momentum $x_F$
and energy $\epsilon_F$ are
\bea
x_F = \frac{p_F}{m_e c} \ , \quad
\epsilon_F= \sqrt{1 +x^2_F}\equiv \tilde{\epsilon}_F+ 1 \ . \label{xr}
\eea
The last equation in eq.~(\ref{hats})  is obtained by the
Sommerfeld expansion (see Appendix C), it agrees with eq.~(6)
in \cite{PC2}. We checked numerically that for our calculations it
is sufficient to take the termodynamic quantities in the Sommerfeld
approximation (SA).

As for the derivative of the electron entropy in respect to the dimensionless WD radius
$x$, let us first consider it in a more transparent SA.
Using the charge neutrality of the plasma (\ref{neut}), the electron
Fermi momentum can be connected to the matter density $\rho$ as follows,
\be
p_F= \hbar\, \bigg( \frac{3\pi^2 \rho}{\mu_u} \bigg)^{1/3}\equiv
D\, \rho^{1/3}\ , \quad
D= \hbar \bigg(\frac{3\pi^2}{\mu_u} \bigg)^{1/3} \ .     \label{de}
\ee
Since $\rho$ is decreasing to the surface of the WD, $p_F$
and $x_F$ are also decreasing, and it follows from
eq.~(\ref{hats}) that $\hat{s}_e$ is increasing.That is,
the specific one-electron  entropy is stratified.
With the help of relations:
\be
x_F = \frac{D}{m_ec}\, \rho^{1/3} \ , \qquad
\epsilon_F= \frac{\sqrt{(m_e c)^2+ D^2 \rho^{2/3}}}{m_ec}   \ , \nn
\ee
the electron entropy  (\ref{hats}) can be transformed to the form, suitable for
calculations,
\be
\hat{s}_e= \frac{\pi^2 k_B T}{c D^2 \rho^{2/3}}\,
\sqrt{(m_e c)^2+ D^2 \rho^{2/3}} \ ,   \label{sseid}
\ee
It is convenient to write the gradient of $\hat{s}_e$ as a product:
\bea
\frac{d\hat{s}_e}{d x} &=&
\left( \rho\, \frac{d \hat{s}_e}{d \rho}\right)\cdot \,
\frac{1}{\rho}\frac{d \rho}{d x}\ , \qquad
\frac{1}{\rho}\frac{d \rho}{d x}< 0 \ , \label{dersh}\\
\rho \frac{d \hat{s}_e}{d \rho} &=&
-\frac{\pi^2  k_B T}{3 c D^2\, \,\rho^{2/3}}\,
\frac{2(m_e c)^2+ D^2 \rho^{2/3}}{\sqrt{(m_e c)^2\,+\,D^2 \rho^{2/3}}}=
- \frac{\pi^2\, \tau}{3}\, \frac{2+ x_F^2}{x_F^2\, \epsilon^2_F}
< 0 \  .   \label{dsehdr}
\eea
Obviously, both terms are dimensionless and negative, hence their
product is dimensionless and positive:
\be
\frac{d\hat{s}_e}{d x} > 0 \ .                 \label{dsedr}
\ee
It should be noted, that our calculations respect the criterion of
the strong degeneracy,
\be
\theta= T/T_F << 1 \ , \quad
{\rm where}\ T_F= \frac{\tilde{E}_F}{k_B} \  ,      \label{SDG}
\ee
and $\tilde{E}_F= E_F- m_ec^2=
c [(m_e c)^2\,+\,p^2_F]^{1/2}- m_e\,c^2=
m_ec^2\, \tilde{\epsilon}_F$\, is the Fermi energy with the rest
mass contribution subtracted.

Due to very good termal conductivity of the WD the temperature $T$ (and hence $\tau=T/T_r$)
are nearly constant inside the WD, with the exception of a thin skip at its surface.
Therefore, in our calculations we consider $\tau$ to be independent of the radius $x$.

We have also checked that the empirical factor
$(1+ \Delta \tilde{\epsilon}/\tilde{\epsilon})^{-1}$
\cite{PC2}, minimizing the
numerical jump of the transition between the fit for $\chi_e < 14$ and
the Sommerfeld expansion for $\chi_e > 14$, did not lead in our calculations to
any sizable effect.

Let us now briefly mention equation for the derivative of the electron entropy following
from the full form of (\ref{hats}). For easier comparison with equations
(\ref{dersh},\ref{dsehdr}) we again use the factorization (\ref{dersh}), where
the 2nd term is now (with the help of (\ref{ne}) and (\ref{neut})):
\bea
\rho \frac{d \hat{s}_e}{d \rho} &=& n_e\, \frac{d \hat{s}_e}{d n_e}=
\frac{n_e}{n'_e} \cdot\, \hat{s}'_e   \ , \label{dsehdrex}\\
n'_e &\equiv&    \frac{d n_e}{d \chi_e}= c_n  \left( I'_{1/2}+ \tau\, I'_{3/2}\, \right)
 \ , \qquad I'_\nu \equiv \frac{d\, I_\nu(\chi_e,\tau)}{d \chi_e } \ , \nn\\
 \hat{s}'_e &\equiv& \frac{d \hat{s}_e}{d \chi_e}=
  \frac{(I_{1/2}+ \tau\, I_{3/2})(5I'_{3/2}+ 4\tau I'_{5/2})-
   (I'_{1/2}+ \tau\, I'_{3/2})(5I_{3/2}+ 4\tau I_{5/2})}{3(I_{1/2}+
 \tau\, I_{3/2})^2} - 1   \ . \nn
\eea
To calculate (\ref{dsehdrex}) one needs derivatives of the Fermi-Dirac integrals $I_\nu(\chi,\tau)$
in respect to $\chi$. It is also not obvious from the general equation that (\ref{dsehdrex})
is negative. Nevertheless, we checked numerically that in our calculations the general
equation (\ref{dsehdrex})  agrees very well with the approximate one (\ref{dsehdr}).

\subsection{The ion entropy}
\label{ientropy}

As for the ions, we consider them in the crystalline phase, in which they
are arranged in  the body-centered cubic (bcc) Coulomb lattice
(see Sect.~3.2.2 of Ref.~\cite{PC3}). In this state, $T < T_m$, where $T_m$
is the melting temperature. For the one-component Coulomb plasma, it is
obtained from the relation,
\be
\Gamma_m= 2.2747 \times 10^5\,\frac{Z^{5/3}}{T_m}\,
\bigg(\rho \frac{Z}{A}\bigg)^{1/3}\,,  \label{Tm}
\ee
where $\Gamma_m= 175\,\pm\,0.4$\,\cite{PC1}.


Beyond the harmonic-lattice approximation (\ref{FM}), the reduced dimensionless one-ion
free energy is given by:
\bea
f_{lat}(\Gamma,\eta) &\equiv& \frac{F_{lat}}{N_{i}\, k_B T}=
 C_M\, \Gamma+  1.5\, u_1\, \eta + f_{th}+ f_{ah}\ .         \label{flat}
\eea
The first three terms describe the harmonic lattice model~\cite{BPY} and
$f_{ah}$ is the anharmonic correction to the Coulomb lattice.
Further, $C_M$ is the Madelung constant~(\ref{cmad}) and \\ $u_1= 0.5113875$.
The parameter $\eta$,
determining the importance of the quantum effects in a strongly
coupled plasma, is \cite{PC3}:
\be
\eta \equiv 7.835 \frac{Z}{A}\cdot \frac{\sqrt{\rho}}{T}\, \times 10^3  \ .   \label{eta}
\ee
The ion coupling parameter $\Gamma\sim 1/T$ is defined by (\ref{Gioncoup}).

For $f_{th}$ we adopt the following fitting formula, used in
the Appendix~B.2 of Ref.~\cite{PC3}:
\be
 f_{th}(\eta) = \sum_{i=1}^3 \ln\big(1 - \mathrm{e}^{-\alpha_i \eta}\big)
- \frac{A(\eta)}{B(\eta)} \ ,     \label{fth}
\ee
where $\alpha_1= 0.932446,\ \alpha_2= 0.334547,\ \alpha_3= 0.265764$ and
\bea
A(\eta) = \sum_{i=1}^7 a_i\, \eta^{m_i} \ , \quad
 B(\eta) = \sum_{i=1}^8 b_i\, \eta^{n_i} \ . \label{ABe}
\eea

\begin{table}
\caption{\label{Tab2} The input data for eqs.~(\ref{ABe}).}
%
\begin{tabular}{|r||c|r|c|r|} \hline
    i&a$_i$&m$_i$&b$_i$&n$_i$  \\ \hline
    1&1.0&0&261.66&0 \\  \hline
    2&0.1839&1&7.07997&2  \\ \hline
    3&0.593 586&2&4.094 84$\times 10^{-2}$&4  \\ \hline
    4&5.4814$\times 10^{-3}$&3&3.973 55$\times 10^{-4}$&5  \\ \hline
    5&5.01813$\times 10^{-4}$&4&5.11148$\times 10^{-5}$&6   \\  \hline
    6&3.9247$\times 10^{-7}$&6&2.19749$\times 10^{-6}$&7    \\  \hline
    7&5.8356$\times10^{-11}$&8&1.866985$\times 10^{-9}$&9   \\  \hline
    8&-&-&2.78772$\times 10^{-13}$&11   \\  \hline
\end{tabular}
\end{table}

For the anharmonic correction of the Coulomb lattice
$f_{ah}$,  we use the anharmonic contribution to the one-ion
entropy from the Sect.~4 of the recent work \cite{BC}.

Using eq.~(\ref{flat}), we calculate the dimensionless one-ion entropy as,
\bea
\hat{s}_i(\Gamma,\eta) &=& - \frac{1}{k_B\, N_i}\,
\frac{ \partial F_{lat}}{\partial T } =
 -f_{lat} - T\, \frac{\partial f_{lat}}{\partial T} =
 -f_{lat} + \Gamma\, \frac{\partial f_{lat}}{\partial \Gamma}
 + \eta\, \frac{\partial f_{lat}}{\partial \eta}  \ ,   \label{hatsi}
\eea
where we used relations
\be
T\, \frac{\partial \Gamma}{\partial T}\, = -\Gamma\ , \quad
T\, \frac{\partial \eta}{\partial T}\, = - \eta \ .   \label{peta}
\ee
It is obvious from (\ref{hatsi}) that the first two terms of (\ref{flat})
(linear in $\Gamma$ and $\eta$, respectively) do not contribute to the entropy
(the corresponding contributions to $F_{lat}$ do not depend on temperature).
From the last part of harmonic contribution, i.e. from
$f_{th}$, we obtain for the
entropy $\hat{s}_i(har)$ two contributions,
corresponding to two terms of (\ref{fth}):
\begin{equation}
\hat{s}_i(har) = \hat{s}_{ths}(\eta)+ \hat{s}_{thr}(\eta) \ ,        \label{sih}
\end{equation}
with
\bea
\label{sbths}
\hat{s}_{ths}(\eta) &=& \sum_{k=1}^3\,\big[ -\ln(1-\mathrm{e}^{-\alpha_k\,\eta})
+\frac{\eta \alpha_k\, \mathrm{e}^{-\alpha_k\,\eta}}
 {1 - \mathrm{e}^{-\alpha_k\,\eta}}\big]\,,  \\
\label{sbthr}
\hat{s}_{thr}(\eta) &=& \frac{A(\eta)}{B(\eta)} -\frac{\tilde{A}(\eta)'}{B}
+A(\eta)\,\frac{\tilde{B}(\eta)'}{B(\eta)^2} \ ,
\eea
where we denote
\be
\tilde{C}(\eta)' \equiv \eta \,\frac{d C(\eta)}{d \eta} \ ,
\quad C= A,\, B \ .  \label{tC}
\ee

The anharmonic part of the one-ion entropy $\hat{s}_i(ah)$ was parameterized in
Ref.~\cite{BC}. From eqs.~(20)\,-\,(24) of this paper one gets:
\bea
\label{siah}
\hat{s}_i(ah) &=& \frac{\tilde{A}^S_1(\eta)}{\Gamma}+
\frac{\tilde{A}^S_2(\eta)}{\Gamma^2}+
\frac{\tilde{A}^S_3(\eta)}{\Gamma^3} \ ,
\qquad \tilde{A}^S_n(\eta)= \eta^n\, A^S_n(\eta) \ , \ n= 1,\, 2,\, 3 \ ,  \\
\label{as1}
\tilde{A}^S_1(\eta)\,&=&\,-2 A_{11}\,
\frac{2 A_{12} \eta^2+ 1}{(1+ A_{12} \eta^2)^2}-
2 A_{13}\,\frac{2 A_{14} \eta^2+ 1}{(1+ A_{14} \eta^2)^2} \ ,   \\
\label{as2}
\tilde{A}^S_2(\eta) &=& \frac{3 A_{2cl}}{2 (1+ A_{21} \eta^4)^{1/4}} \ ,     \\
\label{as3}
\tilde{A}^S_3(\eta) &=& \frac{4 A_{3cl}}{3} \ .
\eea
The parameters entering eqs.~(\ref{as1})-(\ref{as3}) are listed in
Table 1 and in eqs.~(9)-(11)\\ of Ref.~\cite{BC}:
\bea
&& A_{1cl}= 10.2 \ , \quad A_{2cl}= 248 \ , \quad A_{3cl}= 2.03\times 10^5 \ , \nn\\
&& A_{1q}= -0.62/6 \ , \quad A_{2q}= -0.56 \ , \nn\\
&& A_{11}= -10 \ , \quad A_{12}= 6\times 10^{-3} \ , \quad
   A_{13}= -0.2 \ ,\quad A_{14} = 0.2167 \ , \quad
   A_{21}= 2.9624\times 10^{-4} \ . \nn
\eea
%


As for the derivative of the  electron entropy (\ref{dersh}), let us also for
$\hat{s}_i$ factorize:
\bea
\frac{d\hat{s}_i}{d x} &=&
\left( \rho\, \frac{d \hat{s}_i}{d \rho}\right)\cdot \,
\frac{1}{\rho}\frac{d \rho}{d x} \ , \qquad
\frac{1}{\rho}\frac{d \rho}{d x}< 0 \ . \label{dershi}
\eea
Then, using the fact that $\hat{s}_i= \hat{s}_i(\Gamma,\eta)$ and taking into
account the identities:
\be
\rho\, \frac{\partial \Gamma}{\partial \rho}\, = \frac{\Gamma}{3}\ , \quad
\rho\, \frac{\partial \eta}{\partial \rho}\, = \frac{\eta}{2} \ ,
 \label{peg}
\ee
we get for the ion entropy:
\bea
\rho\, \frac{d \hat{s}_i}{d \rho}&=&
\rho\, \frac{\partial \Gamma}{\partial \rho}\,
\frac{\partial \hat{s}_i}{\partial \Gamma}+
\rho\, \frac{\partial \eta}{\partial \rho}\,  \frac{\partial \hat{s}_i}{\partial \eta}=
\frac{\Gamma}{3}\,  \frac{\partial \hat{s}_i}{\partial \Gamma}+
 \frac{\eta}{2}\,  \frac{\partial \hat{s}_i}{\partial \eta} \ . \label{dersip}
\eea

For the harmonic part (\ref{sih}-\ref{sbthr}) it then follows
\begin{equation}
\rho \frac{\partial \hat{s}_i(har)}{\partial\rho}=
\frac{\eta}{2}\,  \frac{\partial \hat{s}_{ths}(\eta)}{\partial \eta}+
\frac{\eta}{2}\,  \frac{\partial \hat{s}_{thr}(\eta)}{\partial \eta}\equiv
ds_{ths}(\eta)+ ds_{thr}(\eta) \ ,        \label{dsih}
\end{equation}
with
\bea
\label{dsths}
ds_{ths}(\eta) &=& -\frac{1}{2}\,\sum_{k=1}^3\,(\eta \alpha_k)^2\,
\frac{\mathrm{e}^{-\alpha_k \eta}}{(1- \mathrm{e}^{-\alpha_k \eta})^2} \ ,   \\
\label{dsthr}
ds_{thr}(\eta) &=& -\frac{A \tilde{B}'^2}{B^3}+
\frac{1}{2 B^2}\,\big(2\tilde{A}' \tilde{B}'+ A \tilde{B}''\big)-
\frac{\tilde{A}''}{2 B} \ ,
\eea
where
\be
\tilde{C}(\eta)'' \equiv \eta^2 \,\frac{d^2 C(\eta)}{d \eta^2} \ , \quad
C= A,\, B \ .  \label{dC}
\ee

In its turn, the derivative of the anharmonic part $\hat{s}_i(ah)$ over $\rho$ is,
\be
\rho \frac{\partial \hat{s}_i(ah)}{\partial\rho}\,=\,
\rho \frac{\partial}{\partial\rho}\bigg[ \frac{\tilde{A}^S_1(\eta)}{\Gamma}\bigg]
 +  \rho \frac{\partial}{\partial \rho}\bigg[\frac{\tilde{A}^S_2(\eta)}{\Gamma^2}\bigg]
 +  \rho \frac{\partial}{\partial \rho}\bigg[\frac{\tilde{A}^S_3(\eta)}{\Gamma^3}\bigg] \ ,      \label{dsiah}
\ee
where we can write similar to (\ref{dersip}),
\bea
\rho \frac{\partial}{\partial \rho}\bigg[\frac{\tilde{A}^S_n(\eta)}{\Gamma^n}\bigg] &=&
\left( \frac{\Gamma}{3}\,  \frac{\partial }{\partial \Gamma}+
\frac{\eta}{2}\,  \frac{\partial }{\partial \eta} \right)\, \frac{\tilde{A}^S_n(\eta)}{\Gamma^n}=
\frac{1}{\Gamma^n}\, \left( - \frac{n}{3}\, \tilde{A}^S_n(\eta)
+ \frac{\eta}{2}\, \frac{d \tilde{A}^S_n(\eta)}{d \eta} \right) \ , \quad n= 1,2,3 \ . \nn
\eea
From the explicit form (\ref{as1})-(\ref{as3}) of factors $\tilde{A}^S_n(\eta)$ one
gets with the help of the relation above:
\bea
\label{das1dg12}
\rho\frac{\partial}{\partial\rho}\bigg[ \frac{\tilde{A}^S_1(\eta)}{\Gamma}\bigg]\,&=&\,
\frac{2}{3\Gamma}\bigg[\frac{A_{11}}{(1\,+\,A_{12}\eta^2)^3}\,
 (8A_{12}^2\eta^4\,+\,3A_{12}\eta^2\,+\,1)\bigg]  \nonumber  \\
 & &\,+\frac{2}{3\Gamma}\bigg[\frac{A_{13}}{(1\,+\,A_{14}\eta^2)^3}\,
(8A_{14}^2\eta^4\,+\,3A_{14}\eta^2\,+\,1)\bigg]\,,  \\
\label{das2dg2}
\rho\frac{\partial}{\partial\rho}\bigg[ \frac{\tilde{A}^S_2(\eta)}{\Gamma^2}\bigg]\,&=&
\,-\frac{A_{2cl}}{4\Gamma^2}\frac{(4\,+\,7 A_{21}\eta^4)}
{(1\,+\,A_{21}\eta^4)^{5/4}}\,,  \\
\label{das3dg3}
\rho\frac{\partial}{\partial\rho}\bigg[ \frac{\tilde{A}^S_3(\eta)}{\Gamma^3}\bigg]\,&=&
\,-\frac{4 A_{3cl}}{3\Gamma^3}\,.
\eea

\begin{figure*}[t]
\begin{tabular}{cc}
 \includegraphics[width=0.332\linewidth]{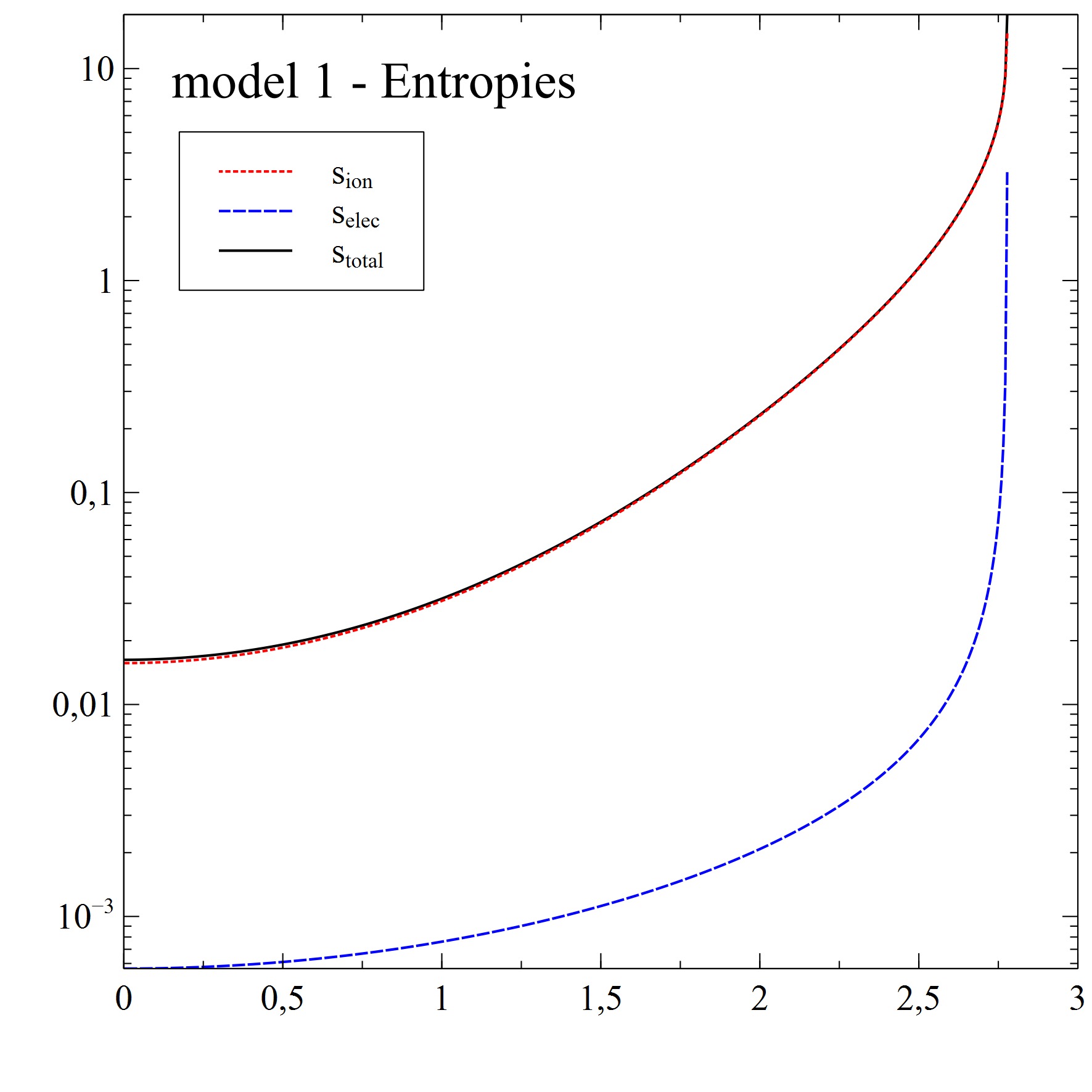}
 &
 \includegraphics[width=0.332\linewidth]{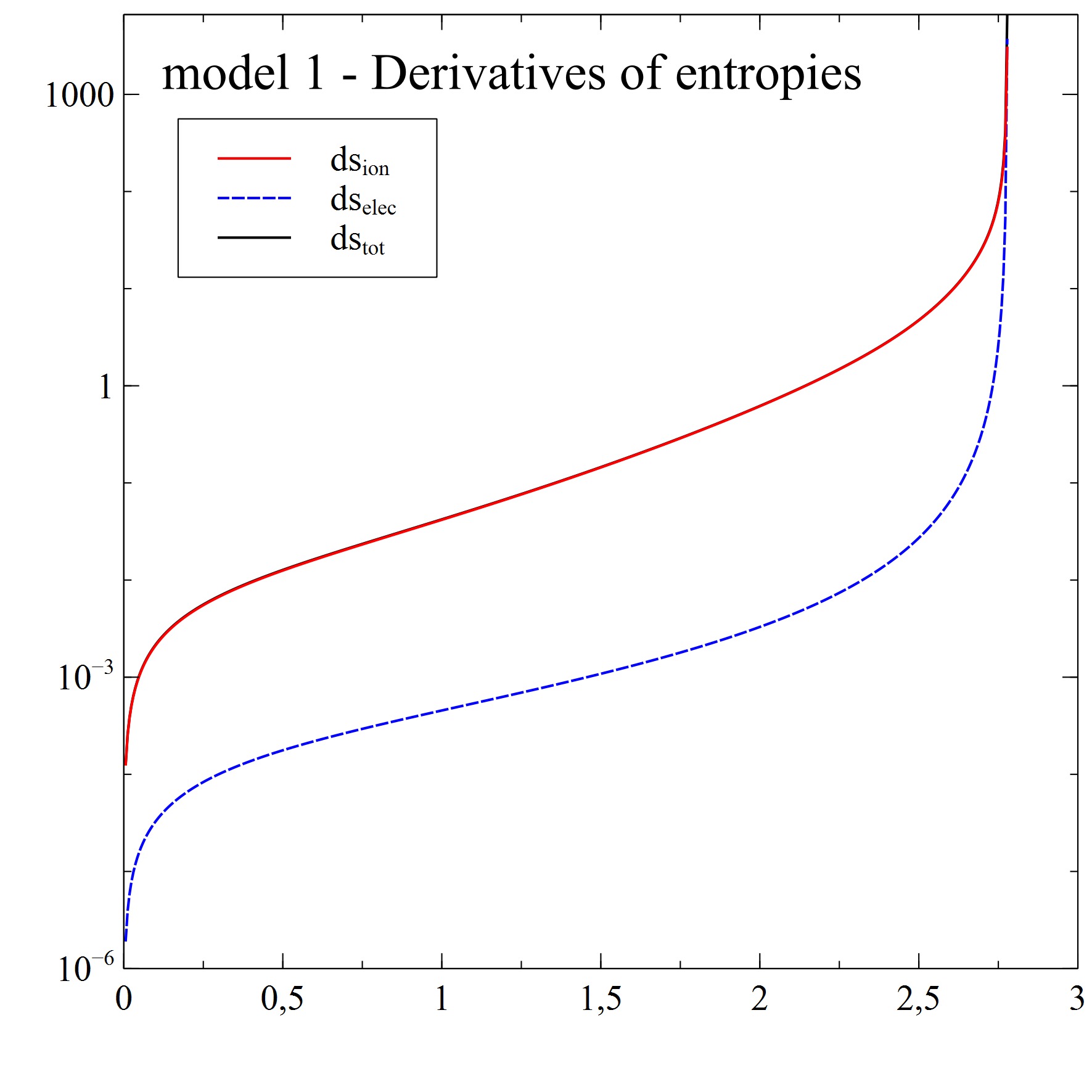}\\
 \includegraphics[width=0.332\linewidth]{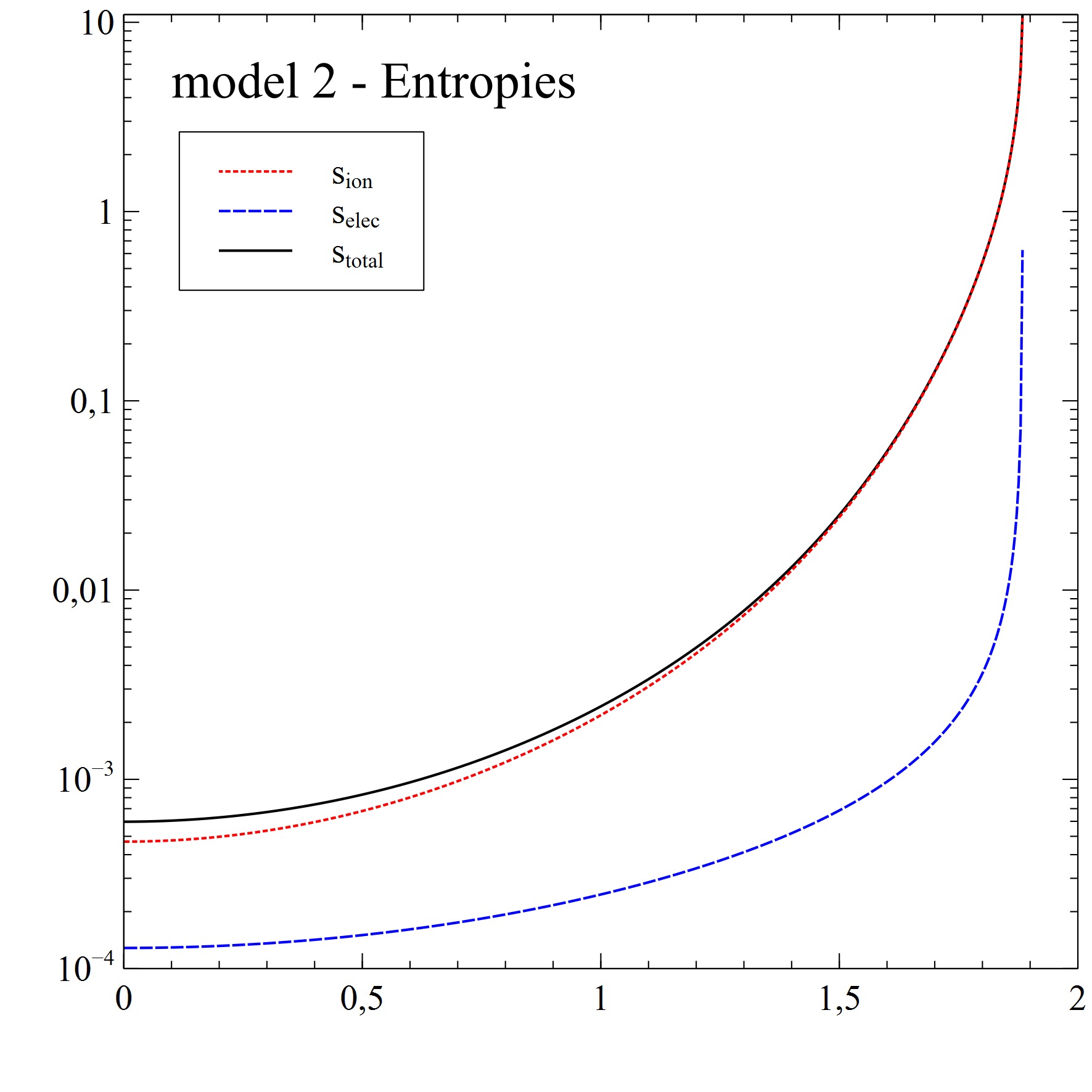}
 &
 \includegraphics[width=0.332\linewidth]{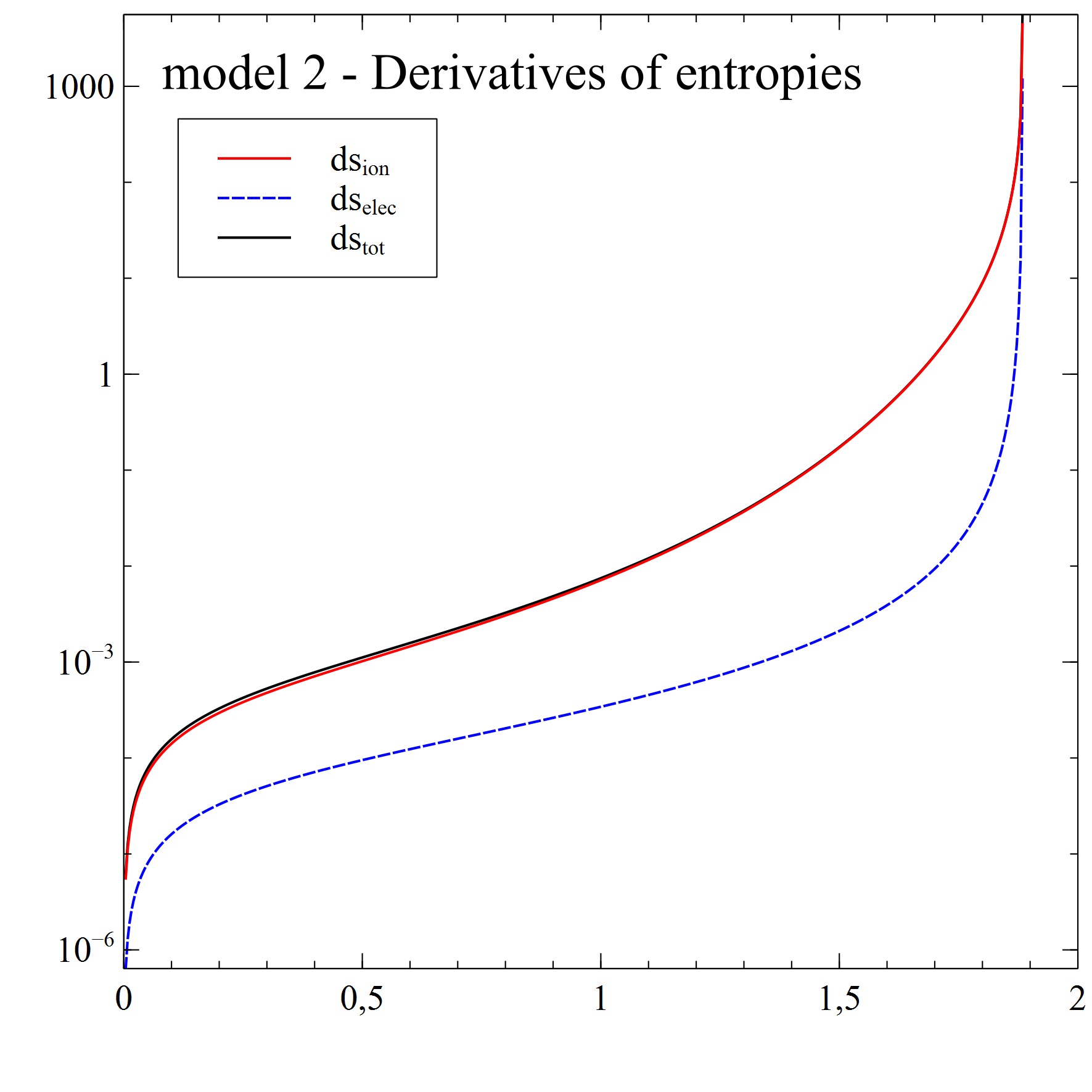}\\
 \includegraphics[width=0.332\linewidth]{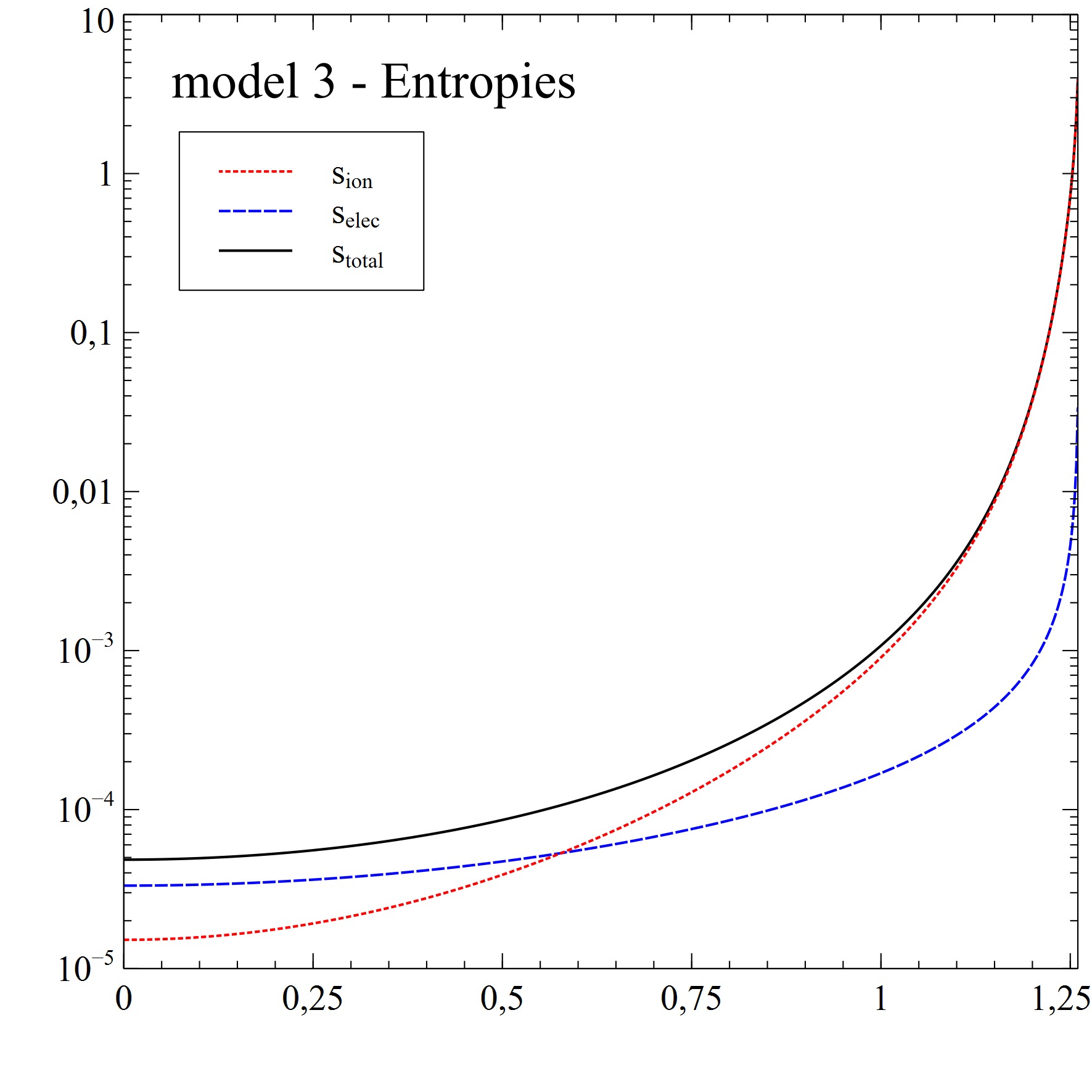}
 &
 \includegraphics[width=0.332\linewidth]{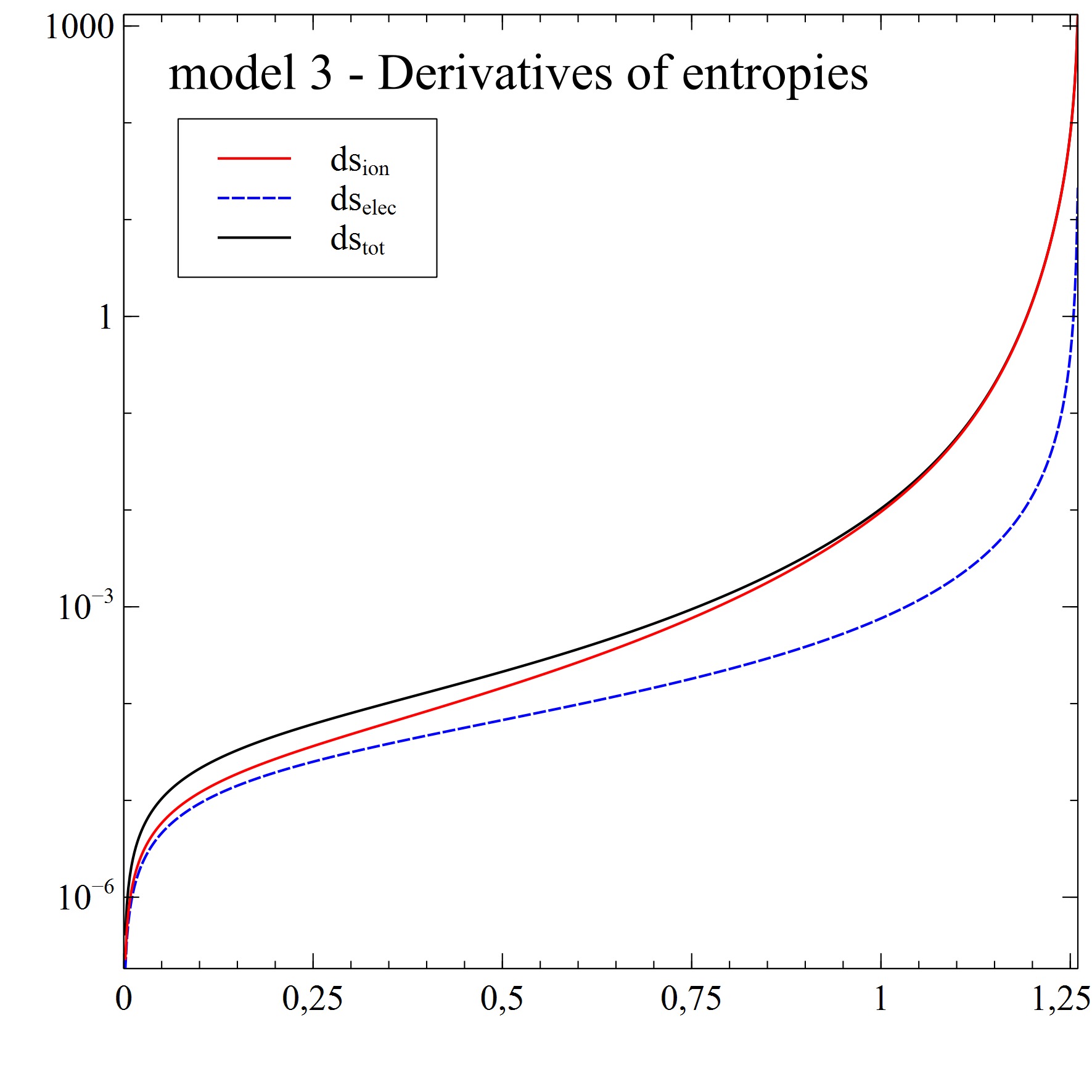}
%
\end{tabular}
\caption{Entropies (left column) and their derivatives (right column)
 for the first three models.\\ On the horizontal axis we plot the
 dimensionless radial distance (\ref{xy}), on the vertical axes we enter
 the dimensional reduced entropy $\hat{s}$ (eq.~(\ref{reducS})) in the left plots and
 $d\hat{s}/d\, x$, i.e.,   derivatives of reduced entropies in respect to
 the dimensionless radius $x$  (eq.~(\ref{xy})) in the right plots.
 As indicated in legends, red dotted curves are ion contributions, blue dashed 
 electron ones and black solid display sums (total values). }
\label{FigL}
\end{figure*}

\begin{figure*}[t]
 \begin{tabular}{cc}
 \includegraphics[width=0.35\linewidth]{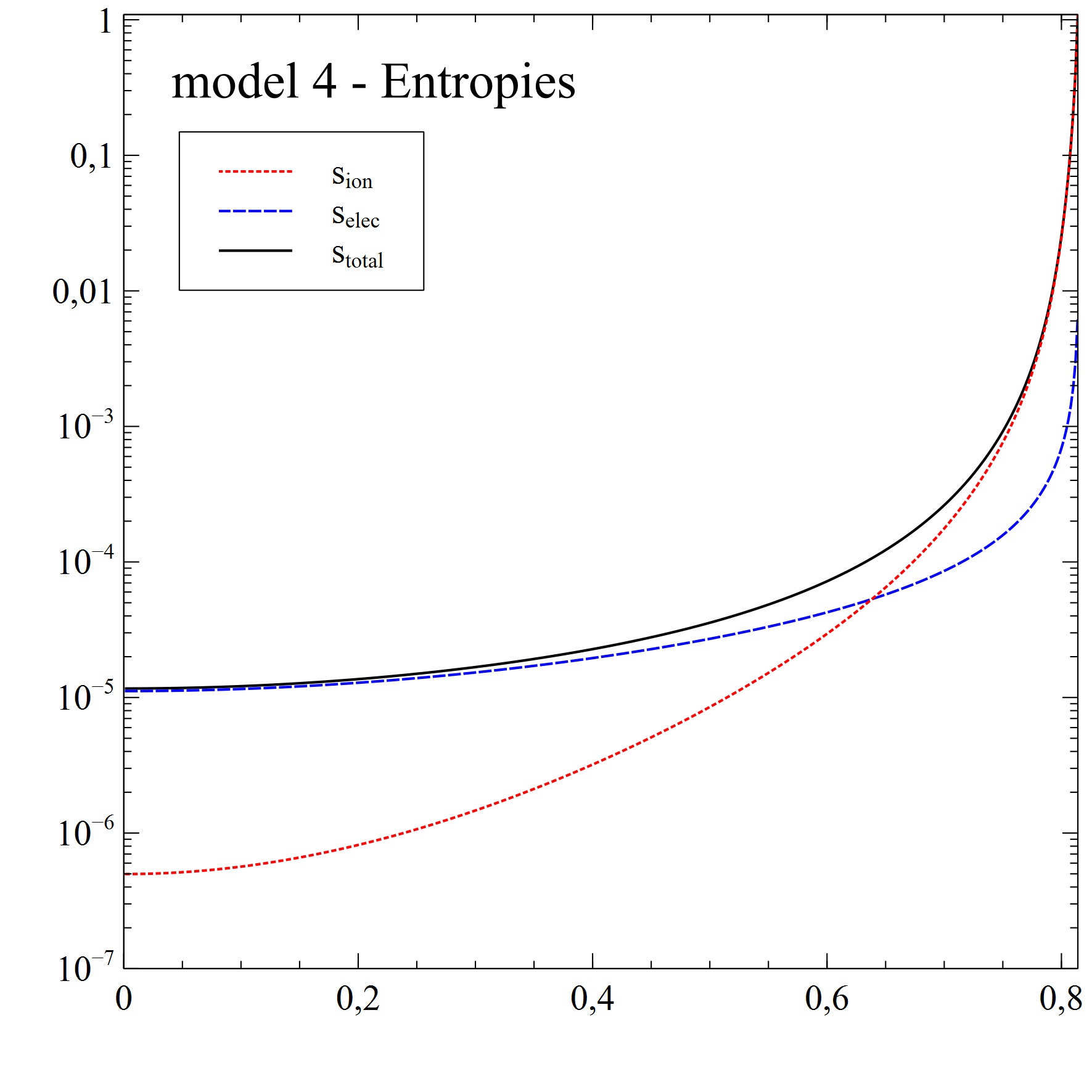}
  &
 \includegraphics[width=0.35\linewidth]{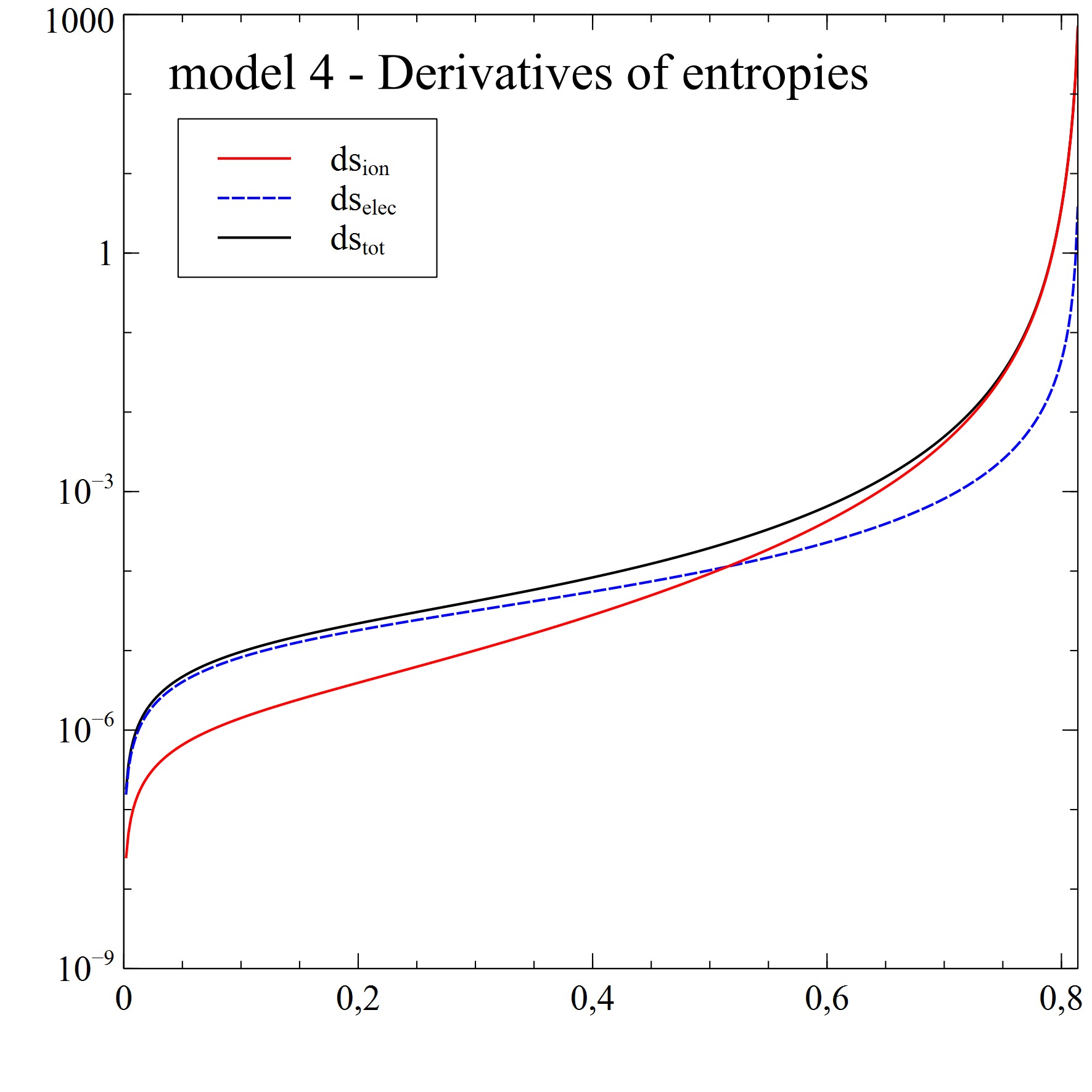}\\
 \includegraphics[width=0.35\linewidth]{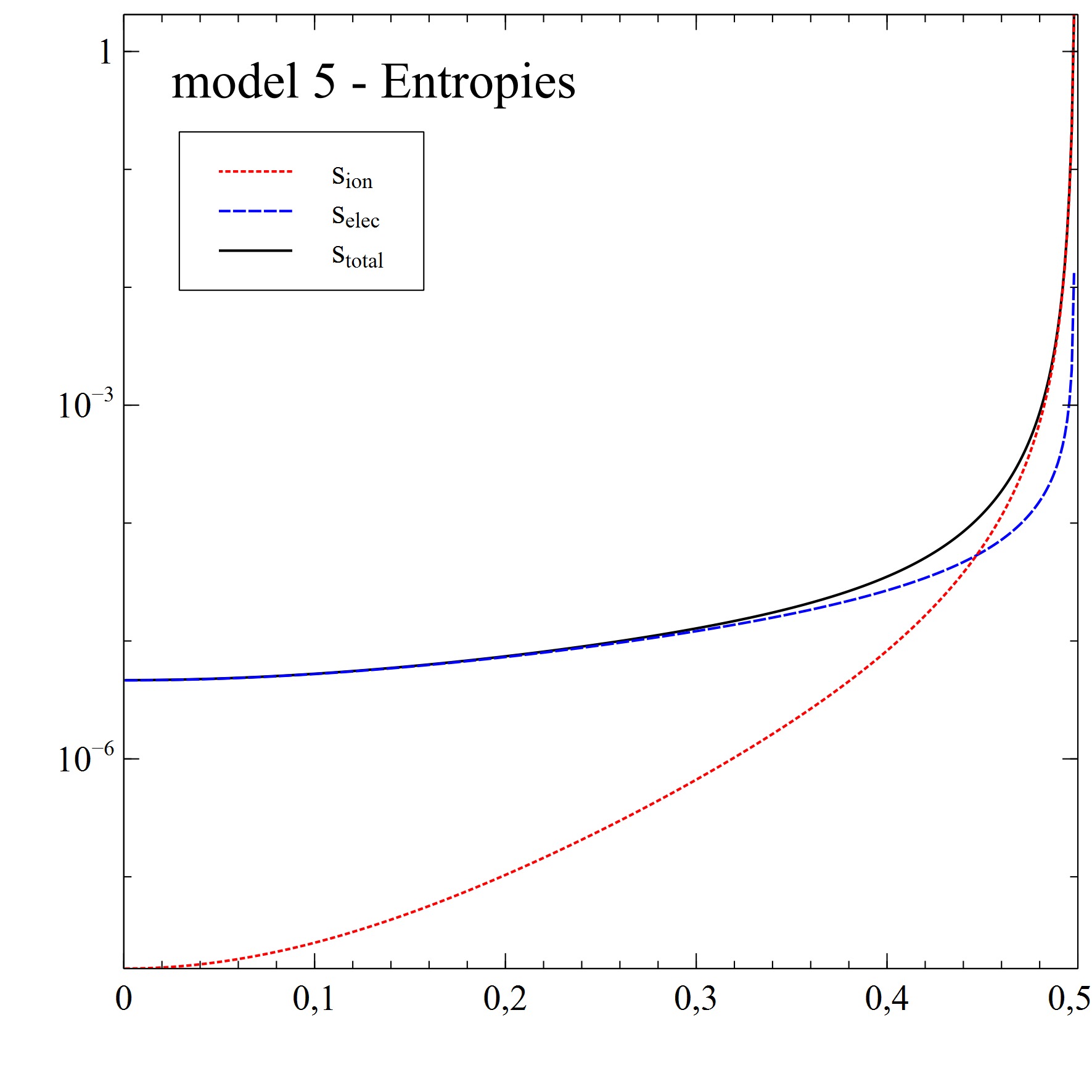}
  &
 \includegraphics[width=0.35\linewidth]{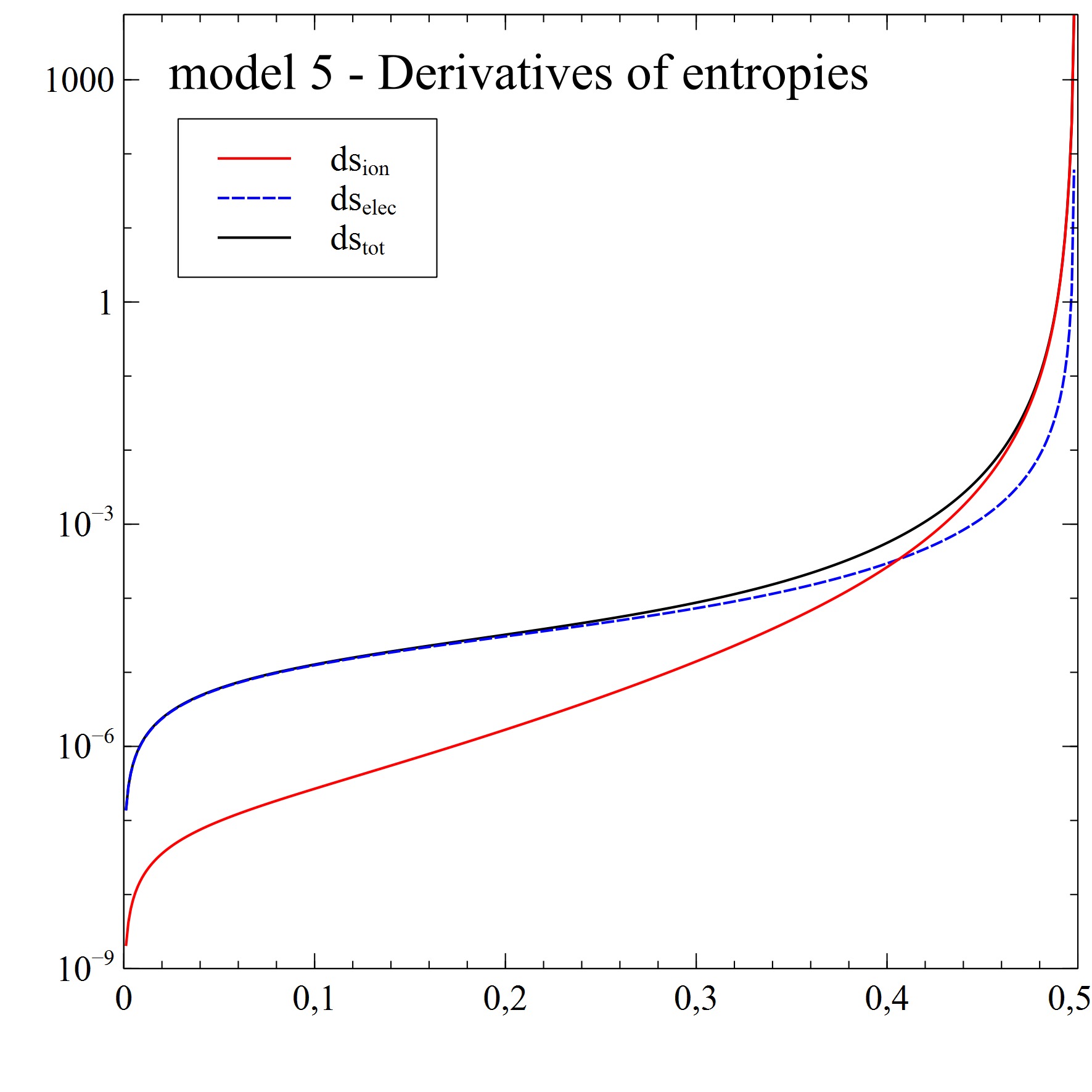}\\
 \includegraphics[width=0.35\linewidth]{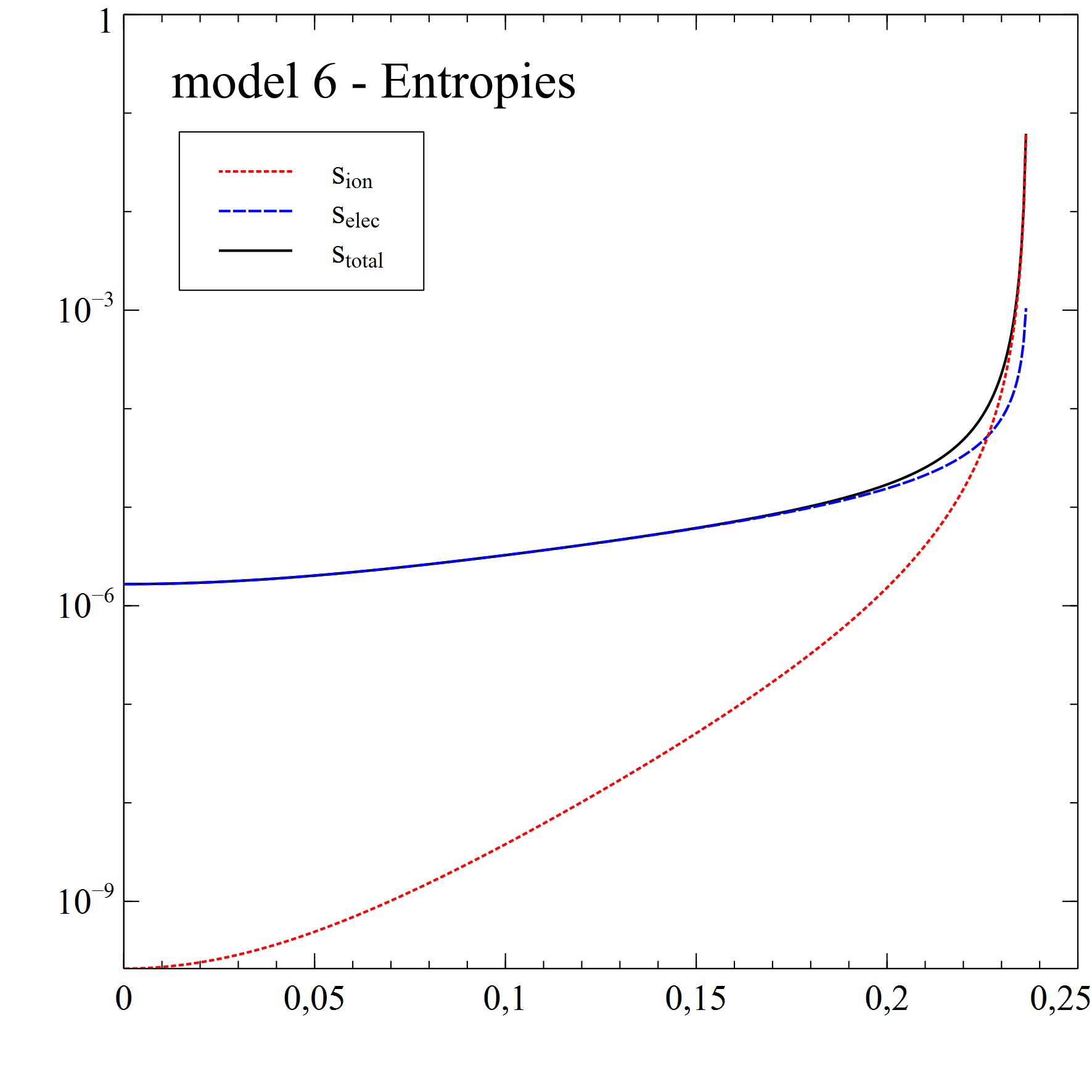}
  &
 \includegraphics[width=0.35\linewidth]{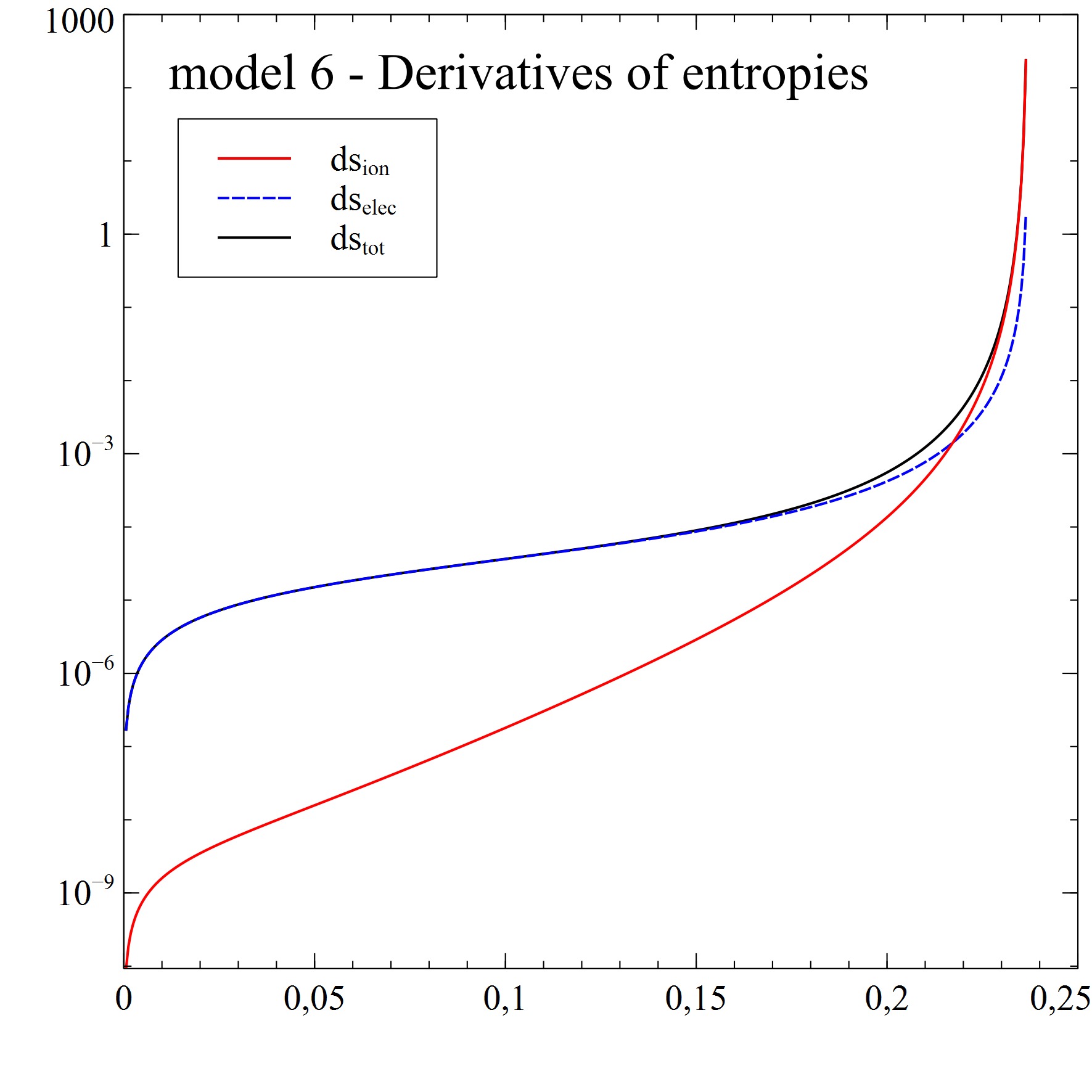}
\end{tabular}
\caption{The same as the previous figure for models 4-6.}
\label{FigH}
\end{figure*}

The electron and ion entropies and their derivatives presented in
this section depend for a given WD (i.e. given $Z,\, A,\, T$ and
$\rho_c$) on the chemical potential $\chi_e$ or on the parameters
$\eta$ and $\Gamma$, resp., which are all determined from the
density $\rho$ obtained by integrating the equation of stability.
This way we  analyzed numerically the WD models 1--6 of
Table~\ref{Tab1}.
The results are plotted in Figs.~\ref{FigL} and \ref{FigH}.\\


\newpage

For the light WDs (see Fig.~\ref{FigL}) the ion entropy prevails, although the
electron contribution becomes gradually important, in particular in the inner parts
of the WDs. The ion contribution also dominates the energy gradient.\\

For heavier WDs  (see Fig.~\ref{FigH}) the electron entropy is more important
almost in the whole star, only close to the surface the ion part prevails.
The same trend is apparent also for entropy gradient.\\

\newpage

As it can be seen from our calculations,
the entropy is positive\footnote{the positivity of the entropy was
noted also in ref.\,\cite{ASJ}} for all models and the entropy gradient satisfies
the condition of the thermodynamical stability of stars Eq.(1.10).


\section{Conclusions}

The frequently used polytropic model  \cite{SC1} of the description
of the WDs has two drawbacks: a) It is of restricted use, because it
is a realistic model of the EoS only in the non-relativistic limit
for $\rho_c << 10^6\, {\rm g/cm^3}$ and
in the  extreme relativistic limit for $\rho_c >> 10^6\, {\rm g/cm^3}$. \\
b) The fluid, described by the polytropic model, is only neutrally stable \cite{JPM}.\\

In this paper, we have shown on a representative set of the carbon
WDs that their description, based on the EoS formulated in the
theory of the magnetized Coulomb plasma in
Refs.~\cite{CP,BPY,PC1,PC2,PC3,ASJ},
satisfies the stability requirement, given by eq.~(\ref{PGE}). As
it is seen in Figs.~\ref{FigL}--\ref{FigH}, both the entropy and its gradient are positive. \\

It would be important to investigate, if this requirement would be
satisfied also in the case of the presence of  the strong magnetic
field. This finding would mean that the existence of strongly
magnetized WDs would be possible.

\section*{Acknowledgments}
We thank Dr.\,A.Y.~Potekhin for the correspondence,
discussions and advices.
The correspondence with Dr.\,N.~Chamel is acknowledged.

\newpage




\bibliographystyle{aipproc}   

\bibliography{at_entropyv2} 

\begin{thebibliography}{41}
\expandafter\ifx\csname natexlab\endcsname\relax\def\natexlab#1{#1}\fi
\providecommand{\enquote}[1]{``#1''}
\expandafter\ifx\csname url\endcsname\relax
  \def\url#1{\texttt{#1}}\fi
\expandafter\ifx\csname urlprefix\endcsname\relax\def\urlprefix{URL }\fi

\bibitem[Chandrasekhar(1939)]{SC1}
Chandrasekhar, S., \emph{An Introduction to the Study of Stellar Structure},
  Dover Publications, inc., University Chicago Press, 1939.

\bibitem[Camenzind(2007)]{MC}
Camenzind, M., \emph{Compact Objects in Astrophysics}, Springer Verlag, Berlin,
  Heidelberg, 2007.

\bibitem[Potekhin(2010)]{AYP}
Potekhin, A.~Y., \emph{Phys. Usp.}, \textbf{53}, 1235 (2010).

\bibitem[Chandrasekhar(1931)]{SC2}
Chandrasekhar, S., \emph{Astrophys. J.}, \textbf{74}, 81 (1931).

\bibitem[Landau(1932)]{LDL}
Landau, L.~D., \emph{Phys. Z. Sowjetunion}, \textbf{1}, 285 (1932).

\bibitem[Shapiro and Teukolsky(1983)]{SLSSAT}
Shapiro, S.~L., and Teukolsky, S.~A., \emph{Black Holes, White Dwarfs, and
  Neutron Stars: The Physics of Compact Objects}, Wiley, New York, 1983.

\bibitem[Mitchell et~al.(2015)]{JPM}
Mitchell, J.~P., Braithwaite, J., Reisenegger, A., Spruit, H., Valvidia, J.~A.,
  and Langer, N., \emph{Mon. Notes. R. Astro. Soc.}, \textbf{447}, 1213 (2015).

\bibitem[Akg\"un et~al.(2013)]{TARMM}
Akg\"un, T., Reisenegger, A., Mastrano, A., and Marchant, P., \emph{Mon. Notes.
  R. Astro. Soc.}, \textbf{433}, 2445 (2013).

\bibitem[Braithwaite(2009)]{JB}
Braithwaite, J., \emph{Mon. Notes. R. Astron. Soc.}, \textbf{397}, 763 (2009).

\bibitem[Spruit(1999)]{HCS}
Spruit, H.~C., \emph{Astron. Astrophys.}, \textbf{349}, 189 (1999).

\bibitem[Bisnovatyi-Kogan(2001)]{CSBK}
Bisnovatyi-Kogan, C.~S., \emph{Stellar Physics 1: Fundamental Concepts and
  Stellar Equilibrium}, Springer, 2001.

\bibitem[Becerra et~al.(2022)]{LB}
Becerra, L., Reisenegger, A., Valvidia, J.~A., and Gusakov, M.~E., \emph{Mon.
  Notes. R. Astron. Soc.}, \textbf{511}, 732 (2022).

\bibitem[Dass and Mukhopadhyay(2013)]{UDM1}
Dass, U., and Mukhopadhyay, B., \emph{Int. J. Mod. Phys. D}, \textbf{22}, 134
  (2013).

\bibitem[Dass and Mukhopadhyay(2014)]{UDM2}
Dass, U., and Mukhopadhyay, B., \emph{J. Cosmol. Astropart. Phys.},
  \textbf{06}, 050 (2014).

\bibitem[Dass and Mukhopadhyay(2015)]{UDM3}
Dass, U., and Mukhopadhyay, B., \emph{J. Cosmol. Astropart. Phys.},
  \textbf{05}, 016 (2015).

\bibitem[Bera and Bhattacharya(2017)]{BB}
Bera, D., and Bhattacharya, D., \emph{Mon. Notes. R. Astron. Soc.},
  \textbf{465}, 4026 (2017).

\bibitem[Chatterjee et~al.(2017)]{DC}
Chatterjee, D., Fantina, A.~F., Chamel, N., Novak, J., and Oertel, M.,
  \emph{Mon. Notes. R. Astron. Soc.}, \textbf{469}, 95 (2017).

\bibitem[Chamel et~al.(2023)]{NCLP}
Chamel, N., Perot, L., Fantina, A.~F., Chatterjee, D., Ghosh, S., Novak, J.,
  and Oertel, M.,  in \emph{The 16th Marcel Grossmann Meeting on General
  Relativity}, edited by R.~Ruffini and G.~Vereshchagin, World Scientific,
  Singapore, 2023, p. 4488 – 4507.

\bibitem[Becerra et~al.(2019)]{LB1}
Becerra, L., Boshkayev, K., Rueda, J.~A., and Ruffini, R., \emph{Mon. Notes. R.
  Astron. Soc.}, \textbf{487}, 812 (2019).

\bibitem[Malheiro et~al.(2012)]{MM}
Malheiro, M., Rueda, J.~A., and Ruffini, R., \emph{Publ. Astron. Soc. Jpn},
  \textbf{64}, 56 (2012).

\bibitem[Ikhsanov and Beskrovnaya(2012)]{IB}
Ikhsanov, N.~R., and Beskrovnaya, N.~G., \emph{Astron. Rep.}, \textbf{56}, 595
  (2012).

\bibitem[Boshkayev et~al.(2013)]{KBLI}
Boshkayev, K., Izzo, L., Rueda, J. A.~H., and Ruffini, R., \emph{Astron.
  Astrophys.}, \textbf{555}, A151 (2013).

\bibitem[Rueda et~al.(2013)]{JAR}
Rueda, J.~A., Boshkayev, K., Izzo, L., Ruffini, R., Loren-Aguilar, P.,
  K\"ulebi, B., Aznar-Sigu\'an, G., and Garcia-Berro, E., \emph{Astropys. J.},
  \textbf{772}, L24 (2013).

\bibitem[Coelho and Malheiro(2014)]{JGC}
Coelho, J.~G., and Malheiro, M., \emph{Publ. Astron. Soc. Jpn.}, \textbf{66},
  14 (2014).

\bibitem[Lobato et~al.(2016)]{RVL}
Lobato, R.~V., Malheiro, M., and Coelho, J.~G., \emph{Int. J. Mod. Phys. D},
  \textbf{25}, 1641025 (2016).

\bibitem[Belyaev et~al.(2015)]{VBB}
Belyaev, V.~B., Ricci, P., \v{S}imkovic, F., Adam, J., Tater, M., and
  Truhl\'{\i}k, E., \emph{Nucl. Phys. A}, \textbf{937}, 17 (2015).

\bibitem[Marsh et~al.(2016)]{TRM}
Marsh, T.~R., G\"ansicke, B.~T., , and H\"ummerich, S., \emph{Nature},
  \textbf{537}, 374 (2016).

\bibitem[Peterson et~al.(2021)]{Dex}
Peterson, J., Dexheimer, V., Negreiros, R., and Castanheira, B.~G.,
  \emph{Astrophys. J.}, \textbf{921}, 1 (2021).

\bibitem[Potekhin and Yakovlev(2012)]{PY}
Potekhin, A.~Y., and Yakovlev, D.~G., \emph{Phys. Rev. C}, \textbf{85}, 030981
  (2012).

\bibitem[Dexheimer et~al.(2014)]{Dex2}
Dexheimer, V., Menezes, D.~P., and Strickland, M., \emph{J. Phys. G: Nucl.
  Part. Phys.}, \textbf{41}, 015203 (2014).

\bibitem[Chabrier and Potekhin(1998)]{CP}
Chabrier, G., and Potekhin, A.~Y., \emph{Phys. Rev. E}, \textbf{58}, 4941
  (1998).

\bibitem[Baiko et~al.(2001)]{BPY}
Baiko, D.~A., Potekhin, A.~Y., and Yakovlev, D.~G., \emph{Phys. Rev. E},
  \textbf{64}, 057402 (2001).

\bibitem[Potekhin and Chabrier(2000)]{PC1}
Potekhin, A.~Y., and Chabrier, G., \emph{Phys. Rev. E}, \textbf{62}, 8554
  (2000).

\bibitem[Potekhin and Chabrier(2010)]{PC2}
Potekhin, A.~Y., and Chabrier, G., \emph{Contib. Plasma Phys.}, \textbf{50}, 82
  (2010).

\bibitem[Potekhin and Chabrier(2013)]{PC3}
Potekhin, A.~Y., and Chabrier, G., \emph{Astron. Astrophys.}, \textbf{550}, A43
  (2013).

\bibitem[Jermyn et~al.(2021)]{ASJ}
Jermyn, A.~S., Schwab, J., Bauer, E., Timmes, F.~X., and Potekhin, A.~Y.,
  \emph{Astrophys. J.}, \textbf{913}, 72 (2021).

\bibitem[Boshkayev(2018)]{KB}
Boshkayev, K., \emph{Astron. Rep.}, \textbf{62(12)}, 847 (2018).

\bibitem[Chamel and Fantina(2015)]{CHFA}
Chamel, N., and Fantina, A.~F., \emph{Phys. Rev. D}, \textbf{92}, 023008
  (2015).

\bibitem[Baiko and Chugunov(2022)]{BC}
Baiko, D.~A., and Chugunov, A.~I., \emph{Mon. Notes R. Astron. Soc.},
  \textbf{510}, 2628 (2022).

\bibitem[Jackson et~al.(2005)]{JTPELP}
Jackson, C.~B., Taruna, J., Pouliot, S.~L., Ellison, B.~W., Lee, D.~D., and
  Piekarewicz, J., \emph{Eur. J. of Phys.}, \textbf{26}, 695 (2005).

\bibitem[Greiner et~al.(1995)]{Grei}
Greiner, W., Neise, L., and Stoeker, H., \emph{Thermodynamics and statistical
  mechanics, Chap. 14}, Springer Verlag, New York, 1995.

\end{thebibliography}


\IfFileExists{\jobname.bbl}{}
{\typeout{}
\typeout{******************************************}
\typeout{** Please run "bibtex \jobname" to optain}
\typeout{** the bibliography and then re-run LaTeX}
\typeout{** twice to fix the references!}
\typeout{******************************************}
\typeout{}
}

\newpage

\appendix

 \section{Scaling And Dimensionless Equations}
\label{appA}

Let us briefly present an alternative derivation of differential
equations describing the WD, starting from the usual Newtonian
formulation of the mechanical stability for the spherical WD (which
is also a starting point in the main text (\ref{EQLIB1})):
\bea
\frac{d\, P_e(r)}{d\, r} &=& - G\, \frac{M(r)\, \rho(r)}{r^2} \ , \nn\\[0.1truecm]
\frac{d\, M(r)}{d\, r} &=& 4\pi\, r^2\, \rho(r) \ , \nn \eea
where $\rho(r)$ is a matter density:
\bea
 \rho(r) &=& m_u\, n_e(r) , \qquad
 \mu_u \equiv \frac{A}{Z}\, m_u \ , \label{muu}
\eea
and $P_e$ is an electron pressure. In the calculations of this paper
we used $A=12, Z=6$ and:
\bea
 m_u= 931.494\, {\rm MeV/c^2} \quad \rightarrow \quad \mu_u= 1\,
862.988\, {\rm MeV/c^2} \ . \nn
\eea
$M(r)$ is a mass contained inside the radius $r$. It appears
convenient to re-write the electron density and pressure in terms of
dimensionless quantities $\tilde{n}_e$ and $\tilde{P}$:
\bea
n_e &=& \rho_0\, \tilde{n}_e \ , \quad 
\rho_0= \frac{1}{3\pi^2\, \lambda^3_e} \ , \quad 
\lambda_e= \frac{\hbar}{m_ec} \simeq 386.164\, {\rm fm}  \ , \label{Lame}\\[0.1truecm]
P_e &=& m_ec^2\, \rho_0\, \tilde{P} \ , \nn
\eea
where $\lambda_e$ is the electron Compton length (its
value is obtained from $m_e\, c^2\simeq 0.511\, $MeV and $\hbar\,
c\simeq 197.33\, {\rm MeV\cdot fm}$).\\
The first equation then reads:
\bea
\frac{d\, \tilde{P}(r)}{d\, r} &=& - G\, \frac{\mu_u}{m_ec^2}\
\frac{M(r)\, \tilde{n}_e(r)}{r^2} \ . \nn
\eea
Next we re-scale also the radius $r$ (as in (\ref{scale})) and the
mass $M(r)$:
\bea
r= a_s\, x \, \qquad  M(R)= M_s\, \tilde{m}(x) \ . \nn
\eea
In terms of $x$ and $\tilde{m}(x)$  the set of differential
equations read:
\bea
\frac{d\, \tilde{P}(x)}{d\, x} &=& - G\, \frac{\mu_u}{m_ec^2}\,
\frac{M_s}{a^2_s} \ \frac{\tilde{m}(x)\, \tilde{n}_e(x)}{x^2} \ , \nn\\[0.1truecm]
\frac{d\, \tilde{m}(x) }{d\, x} &=& 4\pi\, \mu_u\, \rho_0\,
\frac{a^3_s}{M_s}\ x^2\, \tilde{n}_e(x) \ . \nn
\eea
The dimensionless constants appearing on the r.h. sides can be fixed
to our convenience, we adopt a choice (following ref.~\cite{JTPELP}):
\bea
G\, \frac{\mu_u}{m_ec^2}\, \frac{M_s}{a^2_s} &=& \frac{5}{3} \ , \nn\\[0.1truecm]
4\pi\, \mu_u\, \rho_0\, \frac{a^3_s}{M_s} &=& 3 \ , \nn
\eea
from which one gets (cp. (\ref{scale}))
\bea
a_s &=&  \frac{\sqrt{15\pi}}{2 \mu_u}\, \sqrt{\frac{m_ec^2\,
\lambda_e^3}{G}} = \frac{\sqrt{15\pi}}{2}\, \lambda_e\,
\frac{m_{Pl}}{\mu_u}\,
\simeq 8686.26\, {\rm km} \ , \label{myRs}\\[0.1truecm]
M_s &=& a_s\cdot\ \frac{5 m_ec^2}{3\mu m_N G} \simeq
2.646\, M_\odot \ , \label{myMs}
\eea
where $m_{Pl}$ is the Planck mass (with a corresponding value of
the gravitational constant $G$):
\bea
m_{Pl} &\equiv& \sqrt{\frac{\hbar c}{G}} \simeq 1.2209\cdot
10^{22}\, {\rm MeV} \ , \quad G \simeq
6.6742\, \cdot 10^{-8}\, {\rm cm^3/(g\cdot sec^2)} \ . \label{mPLann}
\eea
The set of the dimensionless DE is now:
\bea
\frac{d\, \tilde{P}(x)}{d\, x} &=& - \frac{5}{3}\,
\frac{\tilde{m}(x)\, \tilde{n}_e(x)}{x^2} \ , \nn\\[0.1truecm]
\frac{d\, \tilde{m}(x) }{d\, x} &=& 3\, x^2\, \tilde{n}_e(x) \ . \nn
\eea
To proceed further one has to realize that the quantities
$\tilde{P}$ and $\tilde{n}_e(x)$ are known functions of a
temperature $T$ and electron chemical potential $\mu_e$ (taken here
without the electron rest mass), or more conveniently, of the
dimensionless variables (see the main text):
\bea
\chi_e= \mu_e\, \beta \ , \qquad \tau= \frac{1}{\beta\, m_ec^2} \ ,
\qquad \beta= \frac{1}{k_B\, T} \ . \nn
\eea
An important simplification then follows from an assumption
(employed also in an alternative formulation in the main text) that
to a very good approximation the temperature in the WD is constant,
i.e. $T$ and hence $\tau$ do not depend on $x$ and are fixed by
their initial value. Then all quantities of interest in the WD, in
particular $\tilde{P}$ and $\tilde{n}_e(x)$, depend on the radius
$r$ -- and hence on the dimensionless $x$ -- implicitly just through
a single $x-$dependent function, e.g. $\chi_e(x)$. Thus, we can
write:
\bea
\frac{d\, \tilde{P}(x)}{d\, x} &=& \frac{d\, \tilde{P}(x)}{d\,
\chi_e}\, \frac{d\, \chi_e(x)}{d\, x} \ , \nn
\eea
where the derivative $\tilde{P}(x)/d\, \chi_e$ can be calculated
from the explicit form of $\tilde{P}= \tilde{P}(\chi_e,\tau)$. It is
convenient to consider instead of $\chi_e(x)$ a variable
$\varphi(x)$:
\bea
\varphi(x) &=& \chi_e\, \tau = \frac{\mu_e}{m_ec^2} \ , \nn
\eea
which is a dimensionless electron chemical potential. Its advantage
is that for $T=0$ it just reduces to the dimensionless electron
Fermi energy
\bea
\varphi \ \xrightarrow[T \rightarrow 0]{} \ \frac{\tilde{E}_F}{m_ec^2}
 \equiv \tilde{\varepsilon}_F \ , \label{limphi}
\eea
where $\tilde{E}_F$ (and $\tilde{\varepsilon}_F$) has a contribution of the electron
rest mass subtracted. The resulting set of the DE is then:
\bea
\frac{d\, \varphi(x)}{d\, x} &=& - \frac{5}{3}\,
\frac{\tilde{m}(x)}{x^2}\, g(\varphi)\ , \qquad
g(\varphi)= \frac{\tilde{n}_e}{ \frac{d\, \tilde{P}}{d\, \varphi}}=
\frac{\tau\, \tilde{n}_e(\chi_e,\tau)}{\frac{d\, \tilde{P}(\chi_e,\tau)}{d\, \chi_e}}
\ , \qquad \chi_e(x)= \frac{\varphi(x)}{\tau} \ , \label{alteq1}\\[0.1truecm]
\frac{d\, \tilde{m}(x) }{d\, x} &=& 3\, x^2\, \tilde{n}_e(x) \ .
\label{alteq2}
\eea
As for the initial conditions for these equations:
$\tilde{m}(0)\equiv  \tilde{m}_c= 0$ (and $\tilde{m}(x)$ is an
increasing function of $x$), a value of $ \varphi(0)\equiv \varphi_c
> 0$ is related to the central matter density $\rho_c$ and is
discussed below ($\varphi(x)$ is a decreasing function of $x$).
For the WD the function $g(\varphi)$ is rather close to unity.\\

The set of DE formulated above has several advantages which we are
going to discuss briefly:\\
a) there is a smooth limit of $T \rightarrow 0$. In this limit
$\varphi \rightarrow \tilde{\varepsilon}_F= \sqrt{1+ x^2_F}- 1$, where $x_F$
is a dimensionless electron Fermi momentum. Further, for  the free
electron Fermi gas at $T=0$ it holds:
\bea
\frac{\partial P_e}{\partial E_F}= n_e \ \Rightarrow \
\frac{\partial \tilde{P}}{\partial \varepsilon_F}= \tilde{n}_e \ \Rightarrow
\ g(\varphi) \xrightarrow[T \rightarrow 0]{} 1 \ \ . \nn
\eea
Thus, our set of coupled DEs for $T \rightarrow 0$ smoothly
approaches a set of
\bea
\frac{d\, \tilde{\varepsilon}_F(x)}{d\, x} &=&
- \frac{5}{3}\, \frac{\tilde{m}(x)}{x^2}\ , \nn\\[0.1truecm]
\frac{d\, \tilde{m}(x) }{d\, x} &=& 3\, x^2\, \tilde{n}_e(x) =
3\, x^2\, x^3_F \ , \nn
\eea
which is equivalent to DEs considered in \cite{JTPELP}. This makes
comparisons
of the finite temperature solutions to $T=0$ ones very transparent.\\
b) A numerical solution of  eqs.~(\ref{alteq1}-\ref{alteq2}) is
straightforward: once one specifies the initial conditions, the
equations (complemented by equations for $\tilde{P}(\chi_e,\tau)$
and $\tilde{n}_e(\chi_e,\tau)$) are solved step by step by
appropriate numerical procedure (e.g. 4th order Runge-Kutta) and
there is no need to solve numerically at each step some transcendent
equation (cp to procedure described in a paragraph following
eq.~(\ref{neut})). Moreover, at some point $x_0$ the numerical value
of the decreasing $\varphi(x)$ crosses zero:
\bea
\varphi(x_0)=0 \ \rightarrow \ r_0= a_s\, x_0 \ . \nn
\eea
As in the $T=0$ limit we identify the value of $r_0$ with the
(dimensionless) radius of the WD. The alternative method used in the
main text does not have such a clear criterium for the radius.\\
c) An initial condition for $\varphi(0)= \varphi_c$ is expressed
from the central matter density $\rho_c$ (see eq.~(\ref{xy})). Let
us express
\bea \rho_c= \mu_u\, \tilde{n}_c =  \mu_u\, x^3_{Fc}= \mu_u\,
\tilde{n}_c(\chi_{ec},\tau) \ . \label{phicini} \eea
One can invert this equation numerically to determine $\chi_{ec}$ by
a procedure mentioned below  eq.~(\ref{neut}) (and then to get
$\varphi_c= \tau\, \chi_{ec}$). Let us emphasize that in this
formulation one would have to solve the transcendent equation just
once for the central initial value. But even this is actually not
necessary. At the center of the WD the density is rather high and $T
<< T_F$, hence one can use the Sommerfeld expansion, from which it
is possible to get an algebraic equation for $\varphi_c$ in terms
of $\varphi_{0c}= x^2_{Fc}$ and temperature. We checked that
$\varphi_c$ obtained this way  reproduces very accurately the value
obtained by solving eq.~(\ref{phicini}).\\

In last part of this appendix we briefly present equations for the
WD radius and mass in the LE approximation. There are well known
textbook equations (see e.g. \cite{SLSSAT}) in terms of the central
mass density or one can derive very convenient representations for
dimensionless radii and masses in terms of the central fractional
electron Fermi momentum $x_{Fc}$. To crosscheck numbers in our
Table~\ref{tab-radmas} we used both versions, so we
list below for reference corresponding equations and numerical value.\\

Recall that the radius $R_0$ and the mass of the object described by
the LE equation~(\ref{LEEQ}) are defined by the first zero of
its solution $\theta(\xi_1)= 0$ and by its derivative in
$f(\xi_1)= - \xi^2_1\, \theta'(\xi_1)$
\bea R_0= a\, \xi_1 \ &,& \quad M_0= 4\pi\, f(\xi_1)\, a^3\, \rho_c
\nn \eea
where according to (\ref{XIA}) the LE scaling $a$ is
\bea 
a= \sqrt{\frac{n+1}{4\pi}\, \frac{K}{G}\,
\rho_c^{\frac{1}{n}-1}}= \sqrt{\tilde{K}}\, \rho_c^{\frac{1-n}{2n}}
\ , \qquad \ .  \nn 
\eea

For the non-relativistic case with
\begin{eqnarray}
n= \frac{3}{2} \ , \quad \xi_1^{nr}\simeq  3.65375 \ , \quad
 f(\xi_1^{nr}) \simeq 2.71406 \ ,
\end{eqnarray}
we get:
\bea
K_{nr} &=& \frac{\hbar c\, \lambda_e}{15\pi^2}\,
\left( \frac{3\pi^2}{\mu_N\, m_u} \right)^{5/3} \simeq
3.16119\cdot 10^{12}\, \frac{\rm cm^4}{\rm g^{2/3}\cdot sec^2}   \simeq
\frac{1.00361\cdot 10^{13}}{\mu_N^{5/3}} \, \frac{\rm cm^4}{\rm g^{2/3}\cdot sec^2}
\ , \nn
\eea
which agrees with eq.~(2.3.22) of \cite{SLSSAT}. Then, introducing
\bea
\tilde{K}_{nr} &=& \frac{5}{8\pi}\, \frac{K_{nr}}{G} \simeq
9.42283 \cdot 10^{18}\, {\rm g^{1/3}\cdot cm} \ , \nn
\eea
one gets (substituting $\rho_c[{\rm g/cm^3}]$):
\bea 
R_0[{\rm km}] &=& 10^{-5}\cdot \frac{\sqrt{\tilde{K}_{nr}}\,
\xi_1^{nr}}{\rho_c^{1/6}} \simeq \frac{1.12158\cdot
10^{5}}{\rho_c^{1/6}}= 1.12158\cdot 10^4\, \left( \frac{\mu_N}{2}
\right)^{-5/6}\, \left( \frac{\rho_c}{10^6} \right)^{-1/6}  \ , \nn
\eea
where the last equation agrees with eq.~(3.3.13) of \cite{SLSSAT}.
For the WD mass it follows:
\bea M &=& 4\pi\, \tilde{K}_{nr}^{3/2}\, f(\xi^{nr}_1)\,
\sqrt{\rho_c}
\simeq 9.86510\cdot 10^{29}\, \sqrt{\rho_c} \nn\\[0.1truecm]
&=& 4.95993\cdot 10^{-4} \, M_\odot \, \sqrt{\rho_c}= 0.495993\,
\left( \frac{\mu_N}{2} \right)^{-5/2}\, M_\odot\, \left(
\frac{\rho_c}{10^6} \right)^{1/2} \ , \nn \eea
where we use $M_\odot \simeq 1.98896\cdot 10^{33}\, $g and the
result fairy agrees with eq.~(3.3.14) of \cite{SLSSAT}.

Alternatively, we obtain from (\ref{xr}) and (\ref{de}):
\bea
x_F = \frac{\hbar c}{m_e c^2}\,
\bigg( \frac{3\pi^2 \rho}{\mu_u} \bigg)^{1/3}\, .
\eea
and express in terms of $x_F$ the reduced radius $r_0= R_0/a_s$ and mass
$m_0= M/m_S$, where the scaling factors $a_s$ and $M_s$ are defined
in eqs.~(\ref{myRs},\ref{myMs}). After some algebra one gets:
\bea
r_0 &=& \frac{R_0}{R_s}= \frac{\xi_1^{nr}}{\sqrt{10\, x_{Fc}}}
\simeq \frac{1.15542}{\sqrt{x_{Fc}}}  \ , \label{r0nr}\\[0.2truecm]
m_0 &\equiv& \frac{M}{M_s}=
\frac{3\, f(\xi_1^{nr})}{5\, \sqrt{40}}\, x^{3/2}_{Fc}
\simeq 0.257478\,  x^{3/2}_{Fc} \ , \label{m0nr}\\[0.2truecm]
\frac{M}{M_\odot} &=& \frac{M_s}{M_\odot}\, \cdot m_0=
\frac{\sqrt{6\pi}}{8}\, f(\xi_1^{nr})\,
\frac{m^3_{\rm pl}}{(\mu_N m_u)^2\, M_\odot}\,  x^{3/2}_{Fc}
\nn \\[0.2truecm]
&\simeq& 2.68849\,  m_0
\simeq 0.692227 \,  x^{3/2}_{Fc} \ . \label{mmsunnr}
\eea
These equations are convenient, since $r_0$ and $m_0$ depend only on $x_{Fc}$
and are of natural size. The numbers in the Table~\ref{tab-radmas}
were calculated in both ways, yielding identical results.

For the ultra-relativistic case with
\begin{equation}
n= 3 \ , \quad \xi_1^{ur}\simeq  6.89685 \ , \quad
 f(\xi_1^{ur}) \simeq 2.01824 \ , \nn
\end{equation}
we get:
\bea
K_{ur} &=&  \frac{\hbar c}{12\pi^2}\, \left(\frac{3\pi^2}{\mu_u}
\right)^{4/3}
\simeq 4.93488\cdot 10^{14}\, \frac{\rm cm^3}{\rm g^{1/3}\cdot sec^2}
\simeq \frac{1.24351\cdot 10^{15}}{\mu_N^{4/3}}\,
\frac{\rm cm^3}{\rm g^{1/3}\cdot sec^2} \ , \nn
\eea
which agrees with eq.~(2.3.23) of \cite{SLSSAT}.
We again introduce the auxiliary constant:
\bea
\tilde{K}_{ur} &=& \frac{K_{ur}}{\pi\, G} \equiv
\frac{(3\pi^2)^{1/3}}{4\pi}\, \frac{m^2_{Pl}}{(\mu_N\, m_u)^{4/3}}
 \simeq 2.35357\, \cdot 10^{21} \, {\rm g^{2/3}}  \ . \nn
\eea
Then the radius in km is:
\bea 
R_0[{\rm km}] &=& 10^{-5}\cdot \frac{\sqrt{\tilde{K}_{ur}}\,
\xi_1^{rel}}{\rho_c^{1/3}} \simeq \frac{33.4591}{\rho_c^{1/3}}  =
  0.334591\, \cdot \left( \frac{\mu_N}{2} \right)^{-2/3}\,
\left( \frac{\rho_c}{10^6} \right)^{-1/3}   \ , \nn 
\eea
which is consistent with eq.~(3.3.16) of \cite{SLSSAT}.
The mass is in the ultra-relativistic limit independent of $\rho_c$
and it is known as the Chandrasekhar limit:
\bea
M &=& 4\pi\, \tilde{K}_{ur}^{3/2}\, f(\xi^{rel}_1) \simeq
2.89584\cdot 10^{33}\, {\rm g} \simeq
1.45595\, M_\odot = 1.45595\, \left(\frac{\mu_N}{2} \right)^{-2}\, M_\odot  \ . \nn
\eea
Alternatively, we can calculate $R_0$ and $M/M_\odot$ in terms of the dimensionless:
\bea
r_0 &=& \frac{\xi_1^{rel}}{\sqrt{5}\, x_{Fc}}
\simeq \frac{3.084370}{x_{Fc}} \ , \quad
R_0= R_s\cdot r_0 \ , \label{r0rel}\\[0.2truecm]
m_0 &\equiv& \frac{M}{M_s}=
\frac{3\, f(\xi_1^{rel})}{5\, \sqrt{5}} \simeq 0.541550 \ , \label{m0rel}
\eea
from which one gets the same result as above:
\bea
\frac{M}{M_\odot} &=& \frac{M_s}{M_\odot}\, \cdot m_0=
\frac{\sqrt{3\pi}}{2}\, f(\xi_1^{ur})\,
\frac{m^3_{pl}}{(\mu_N m_u)^2\, M_\odot}
\simeq  1.45595 \ . \label{mmsunrel}
\eea

\newpage

\section{Calculations of functions $f_1(y)$ and $f_2(y)$}
\label{appB}

In accord with eq.~(23) of \cite{PC3}, the derivative of the electron
pressure $P^{(e)}_{id} $ over $y$ is
\be 
\frac{ \partial P^{(e)}_{id} }{\partial y}= \rho_c\, \frac{
\partial P^{(e)}_{id} }{\partial \rho}=
\rho_c\,\frac{n_e}{\rho}\, \bigg(\frac{ \partial P^{(e)}_{id}} {
\partial \chi_e}\bigg)_T \bigg/ \bigg(\frac{\partial n_e}{\partial
\chi_e}\bigg)_T \ ,   \label{FDP} 
\ee
and, similarly, the second derivative is
\be 
\frac{ \partial ^2 P^{(e)}_{id} }{\partial y^2}=
\rho_c^2\,\frac{n_e^2}{\rho^2}\, \bigg[\,\bigg(\frac{\partial^2
P^{(e)}_{id} } {\partial \chi_e^2}\bigg)- \bigg(\frac{ \partial
P^{(e)}_{id}} {\partial \chi_e }\bigg)\, \bigg(\frac{ \partial^2
n_e} {\partial \chi_e^2 }\bigg)\, \bigg/\,\bigg(\frac{\partial n_e}
{\partial \chi_e}\bigg)\, \bigg]_T\, \bigg/ \bigg(\frac{\partial
n_e}{\partial \chi_e}\bigg)^2_T \ .    \label{SDP} 
\ee
So one should calculate $n_e$ and  the derivatives
of $ P^{(e)}_{id} $ and $ n_e$ over $ \chi_e$ in
terms of the Fermi-Dirac integrals $I_{\nu}(\chi_e,\,\tau)$.
Identifying
\bea
\label{Ik} I_{k+1/2}(\chi_e,\,\tau) &\equiv& Wk \ ,
\\
\label{dIk}
\frac{\partial I_{k+1/2}(\chi_e,\,\tau )}{\partial \chi_e} &\equiv& WkDX \ ,
\\
\label{ddik}
\frac{\partial^2 I_{k+1/2}(\chi_e,\,\tau )}{\partial \chi_e^2 }
&\equiv& WkDXX \ ,
\eea
one can write the  $n_e$ and  the derivatives of $ P^{(e)}_{id} $ and
$ n_e$ over $ \chi_e$ in terms of $Wk,\, WkDX$ and $WkDXX$.
We calculated these quantities using the program  BLIN9~\cite{PC3}
and expressed in terms of them:

\bea
\label{neb}
n_e &=& c_n\, [\,W0+ \tau\,W1\,] \equiv c_n\, Z_5 \ ,
\\
\label{dneb}
\frac{\partial n_e}{\partial \chi_e} &=&
c_n\, [\,W0DX+ \tau\,W1DX\,] \equiv  c_n\, Z_3 \ ,
\\
\label{dpeb}
\frac{\partial P^{(e)}_{id}}{\partial \chi_e} &=&
c_p\, [\,W1DX+ \frac{\tau}{2}\cdot \,W2DX\,] \equiv c_p\,Z_4 \ ,
\\
\label{ddneb}
\frac{\partial^2 n_e}{\partial \chi_e^2} &=&
c_n\,[\,W0DXX+ \tau\,W1DXX\,] \equiv c_n\, Z_2 \ ,
\\
\label{ddpeb}
\frac{\partial^2 P^{(e)}_{id} } {\partial \chi_e^2 } &=&
 c_p\, [\,W1DXX+ \frac{\tau}{2}\cdot\, W2DXX\,] \equiv c_p\,Z_1 \ .
\eea
In terms of $Z_i$ we get eq.~(\ref{FDP}) in the form
\be
\frac{\partial P^{(e)}_{id} }{\partial y}=
\frac{c_p}{y} \,\frac{Z_4\,Z_5}{Z_3}\ ,   \label{FDPF}
\ee
and eq.~(\ref{SDP})  becomes
\be
\frac{ \partial ^2 P^{(e)}_{id} }{\partial y^2}=
\frac{c_p}{y^2}\,\frac{ Z_5^2}{ Z_3^3}\, (Z_1\, Z_3 - Z_2 \,Z_4 ) \ . \label{SDPF}
\ee
Finally,
\be
f_1(y) = \frac{1}{y}\,\bigg[\frac{Z_5}{Z_3^2\,Z_4}\,(Z_1\, Z_3 - Z_2 \,Z_4
-1\,\bigg] \ , \quad
f_2(y)= \frac{c_p}{y}\,\frac{Z_4\,Z_5}{Z_3} \ . \label{f1yf2yf}
\ee

\newpage

\section{The Sommerfeld expansion}
\label{appC}

In this Appendix we briefly describe how to decompose the thermodynamical
quantities for free electrons into series in powers of $k_B T/\tilde{E}_F$,
where $\tilde{E}_F= \mu_e(T=0)$  is the Fermi energy with the rest mass
contribution subtracted.  We will start from simpler non-relativistic dynamics and
later extend the results to a general case.

In the non-relativistic approximation we can write for electron
density, momentum and energy density in a conveniently normalized form:
\bea
\left(\frac{n_e(T)}{\rho_0} \right)_{nr} &=&
3\sqrt{2}\, \tau^{3/2}\, I_{1/2}(\chi_e)   \ , \label{rhoexnrT}\\[0.2truecm]
\left(\frac{P(T)}{\rho_0\, m_ec^2} \right)_{nr} &=&
2\sqrt{2}\, \tau^{5/2}\, I_{3/2}\chi_e)   \ , \label{PexnrT}\\[0.2truecm]
\left(\frac{{\cal E}_e(T)}{{\cal E}_0} \right)_{nr} &=&
\sqrt{2}\, \tau^{5/2}\, I_{3/2}(\chi_e) =
\frac{1}{2} \left(\frac{P(T)}{\rho_0\, m_ec^2} \right)_{nr}
\ ,  \quad  {\cal E}_0= 3\rho_0\, m_ec^2\ , \label{EexnrT}
\eea
where the non-relativistic Fermi-Dirac integrals are here defined as:
\bea
I_{\nu}(\chi) =  \int\limits_0^\infty\,
\frac{u^{\nu}}{\mathrm{e}^{u-\chi}+1}\, du \ ,   \label{FDnonrel}
\eea
Substituting into the 1st line of (\ref{hats}) one gets the reduced entropy in the
non-relativistic limit:
\bea
s_{e, nr} &=& \frac{1}{k_B T}\,
\left( \frac{5\, m_ec^2\, \tau\, I_{3/2}(\chi_e)}{3\,  I_{1/2}(\chi_e)}
- \mu_e  \right)=
\frac{5\, I_{3/2}(\chi_e)}{3\,  I_{1/2}(\chi_e)} - \chi_e \ . \nn
\eea
This non-relativistic limit follows also from the general results (\ref{hats})
with the help of the relation:
\begin{equation}
I_\nu(\chi,\tau) \  \xrightarrow[\tau\rightarrow 0]{} \  I_{\nu}(\chi) \ . \nn
\end{equation}
An opposite ultra-relativistic limit is obtained from
\bea
I_\nu(\chi,\tau) \  \xrightarrow[\tau\rightarrow \infty]{} \
\sqrt{\frac{\tau}{2}}\,  I_{\nu+1/2}(\chi) \ . \nn
\eea
The number density, momentum and energy density in the ultra-relativistic limit are:
\bea
\left(\frac{n_e(T)}{\rho_0} \right)_{ur} &=&
3 \tau^3\, I_2(\chi_e)   \ , \label{rhoexurT}\\[0.2truecm]
\left(\frac{P(T)}{\rho_0\, m_ec^2} \right)_{ur} &=&
\left(\frac{{\cal E}_e(T)}{{\cal E}_0} \right)_{ur}=
\tau^4\, I_3(\chi_e)  \ . \label{PexurT}
\eea
Equations above define the densities of the electron number,
pressure and kinetic energy as functions of $\chi_e= \beta\, \mu_e$
and $\tau= k_B T/m_ec^2$,
i.e., functions of the chemical potential $\mu_e$ (not yet determined) and of
the  temperature $T$ (or of $\beta= k_B T$).  According to ref.~\cite{PC3} the chemical potential
$\mu_e(V,T)$ is obtained by (numerically) inverting equation for the density
(\ref{ne}) (in its exact or non/ultra-relativistic forms).
Assuming the fixed number of electrons $N_e$ (meaning that the electron
density  $n_e= N_e/V$ explicitly depends only on volume $V$),
one gets the chemical potential from  the condition $n_e(T)=n_e(0)= N_e/V$,
where $n_e(0)$ is known function of the Fermi energy. At the end we get the
chemical potential dependent on the temperature $T$ and the Fermi energy,
which is its value for zero temperature:
\bea
 \mu_0&\equiv& \mu_e(T=0)= \tilde{E}_{F}= m_e\, c^2\,
\tilde{\epsilon}_{F}   \ , \label{mu02nr}
\eea
where the chemical potential and energies do not include the rest mass contributions.

For a simple non-relativistic case, for which $ \mu_0= \tilde{E}_{Fnr}/(m_e\, c^2)=
\tilde{\epsilon}_{Fnr}= x_F^2/2$\\ with $x_F=p_F/(m_ec)$,
the l.h.s. of (\ref{rhoexnrT}) at $T=0$ reads:
\bea
\left(\frac{n_e(0)}{\rho_0} \right)_{nr} &=& x^3_F=
\left( 2\, \tilde{\epsilon}_{Fnr}\right)^{3/2}=
\left( 2\, \frac{\mu_0}{m_ec^2}\right)^{3/2}=
2\, \sqrt{2}\, \left( \frac{\mu_0}{m_ec^2} \right)^{3/2}  \ . \nn
\eea
Equating this to the r.h.s and substituting for $\tau$ yields:
\bea
2\, \sqrt{2}\, \left( \frac{\mu_0}{m_ec^2} \right)^{3/2} =
3\, \sqrt{2}\, \left( \frac{k_B T}{m_ec^2} \right)^{3/2}\,
I_{1/2}(\chi_e) \ . \nn
\eea
which simplifies to:
\bea
I_{1/2}(\chi_e)= \frac{2}{3}\, \left(\frac{\tilde{\mu}_0}{k_B T}\right)^{3/2}
= \frac{2}{3}\, \left(\frac{\tilde{E}_{Fnr}}{k_B T}\right)^{3/2}
= \frac{2}{3}\, \left(\frac{T_F}{T}\right)^{3/2}  \ . \label{munrcond2}
\eea
Denoting by $X_{1/2}$ the inverse function to $I_{1/2}(\chi_e)$ one gets
a solution for $\chi_e$:
\bea
\chi_e&\equiv& \frac{\mu}{k_B T} =
X_{1/2} \left( \frac{2}{3}\, \left(\frac{T_F}{T} \right)^{3/2}\,  \right)
\ , \nn
\eea
which is just eq.~(17) of ref.~\cite{CP} in our notations. In general,
a similar connection is obtained from $n_e$ given by (\ref{ne}), in the
ultra-relativistic limit from (\ref{rhoexurT}).

The relations above are valid for arbitrary temperature, now we will deal
with a low temperature expansion.
The non-relativistic Fermi-Dirac integrals (\ref{FDnonrel}) can be
for small temperatures (i.e. large inverse temperatures $\beta= 1/k_BT$ and
hence also $\chi_e= \beta\, \mu_e$) approximated by a power series:
\bea
I_\nu(\chi) &\simeq&  \frac{\chi^{\nu+1}}{\nu+1}\, \Big(1
+ \frac{\pi^2}{6}\, \frac{(\nu+1)\nu}{\chi^2} +
\frac{7\, \pi^4}{360}\,  \frac{(\nu+1)\nu(\nu-1)(\nu-2)}{\chi^4}
+ \dots  \Big) \ . \label{FDsimsom}
\eea
\noindent
For the non-relativistic dynamics one needs $I_\nu(\chi)$ with $\nu= 1/2$
and $\nu= 3/2$:
\bea
I_{1/2}(\chi) &\simeq& \frac{2}{3}\, \chi^{3/2}\,
\Big( 1+ \frac{\pi^2}{8\, \chi^2} + \frac{7\, \pi^4}{640\, \chi^4}
+ \dots \Big) \ , \label{SommnrF12}\\[0.2truecm]
I_{3/2}(\chi) &\simeq& \frac{2}{5}\, \chi^{5/2}\,
\Big( 1+ \frac{5\, \pi^2}{8\, \chi^2} - \frac{7\, \pi^4}{384\, \chi^4}
+ \dots \Big) \ . \label{SommnrF32}
\eea
Substituting these relations into (\ref{rhoexnrT})-(\ref{EexnrT})
and using $\tau\, \chi_e= \mu_e/(m_ec^2)$
we get:
\bea
\frac{1}{3\sqrt{2}}\left(\frac{n_e (T)}{\rho_0} \right)_{nr} &=&
\tau^{3/2}\, I_{1/2}(\chi_e)  \simeq
\frac{2}{3} \left( \frac{\mu_e}{m_ec^2} \right)^{3/2}\,
\left( 1+ \frac{\pi^2}{8 \chi_e^2}+ \frac{7 \pi^4}{640 \chi_e^4}+ \dots \right)
\ , \label{rhoexnrTS}\\[0.3truecm]
\frac{1}{2^{3/2}} \left(\frac{P(T)}{\rho_0\, m_ec^2} \right)_{nr} &=&
\frac{1}{\sqrt{2}}
\left(\frac{{\cal E}_e(T)}{{\cal E}_0} \right)_{nr} =
\tau^{5/2}\, I_{3/2}(\chi_e) \nn\\[0.1truecm]
&\simeq&   \frac{2}{5} \left( \frac{\mu_e}{m_ec^2} \right)^{5/2}\,
\left( 1+ \frac{5 \pi^2}{8 \chi_e^2}- \frac{7 \pi^4}{384 \chi_e^4}+ \dots \right)
\ . \label{PexnrTS}
\eea
These are formal power series in terms of powers of the so far
unknown $1/\chi_e^2\, $. Recall that $\chi_e$ depends on temperature and
on the chemical potential. As discussed above, $\mu_e$
is determined from the condition $n_e(T)= n_{e}(0)$ (but now with
$n_e(T)$ decomposed into the power series above).
Substituting $1/\chi_e^2= (kT)^2/ \mu_e^2$ (and using
$\mu_0= \tilde{E}_{Fnr}$ for the non-relativistic Fermi energy) leads to:
\bea
 \mu_0 &=& \mu_e\, \left( 1
 + \frac{1}{8} \frac{(\pi k_B T)^2}{\mu_e^2}
+ \frac{7}{640} \frac{(\pi k_B T)^4}{\mu_e  ^4}+ \dots \right)^{2/3}
\ . \label{tilmunr}
\eea
This relation can be perturbatively inverted by assuming the power series for
$\mu_e$ (in powers of $(k_B T/\mu_0)^2$):
\bea
\mu_e &=&  \mu_0\, \left( 1- \frac{\pi^2}{12}\frac{(k_B T)^2}{ \mu_0^2}
- \frac{\pi^4}{80}\frac{(k_ B T)^4}{ \mu_0^4}+ \dots \right) \ . \label{tilmunrT}
\eea
Substituting (\ref{tilmunrT}) into the r.h.s. of  and
(\ref{PexnrTS}) and making the Taylor decomposition in powers of $(k_B T)^2$
yields the following non-relativistic equations for the observables:
\bea
P_{nr}(T) &=&   m_ec^2\, \rho_0\ \frac{x^5_F}{5}\,
\left(1 + \frac{5\pi^2}{12} \frac{(k_B T)^2}{\mu^2_0}
- \frac{\pi^4}{16} \frac{(k_B T)^4}{\mu^4_0} + \dots \right)
\ , \label{PexnrSomm} \\[0.3truecm]
{\cal E}_{e,nr}(T) &=& \frac{3}{2}\, P_{nr}(T) \ . \label{EexnrSomm}
\eea
These results are consistent with the equations, presented in Ref.~\cite{Grei}
(where slightly different notation is used). \\

For the ultra-relativistic dynamics one proceeds in a similar way. For sake of briefness,
we cite just the final results (using $\mu_0/(m_ec^2)=\epsilon_{Fur}=x_F$):
\bea
\mu_e &=& \mu_0\, \left( 1 - \frac{\pi^2}{3}
\frac{(kT)^2}{\mu_0^2}  + O(T^6) \right) \ . \label{tilmuurT}\\[.2truecm]
n_{e,ur}(T) &=&  n_{e,ur}(0)=
\rho_0\, \left( \frac{\mu_0}{m_ec^2}\right)^3=
\rho_0\, \epsilon^3_{Fur} = \rho_0\, x_F^3
= \frac{N_e}{V} \ , \label{rhoexurSomm}\\[.2truecm]
P_{ur}(T) &=& \rho_0\, m_ec^2\, \frac{x^4_F}{4}\,
\left(1+ \frac{2\pi^2}{3}\frac{(k_B T)^2}{\mu^2_0}
- \frac{\pi^4}{5}\frac{(k_B T)^4}{\mu^4_0}+ \dots \right)
\ , \label{PexurSomm}\\[.2truecm]
{\cal E}_{e,ur}(T) &=& 3\, P_{ur}(T) \ . \label{EexurSomm}
\eea

When the non-relativistic or ultra-relativistic limits cannot be applied,
we start from normalized equations (\ref{ne}-\ref{Ueid}):
\bea
\frac{n_e(T)}{\rho_0} &=&  3\sqrt{2}\, \tau^{3/2}\, \left[
I_{1/2}(\chi_e,\tau)+ \tau\, I_{3/2}(\chi_e,\tau) \right]
\ , \quad \rho_0= \frac{1}{3\pi^2\, \lambda^3_e} \ , \label{rhoexT}\\[0.2truecm]
\frac{P(T)}{\rho_0\, m_ec^2} &=&   2\sqrt{2}\, \tau^{5/2}\,
\left[ I_{3/2}(\chi_e,\tau)+ \frac{\tau}{2}\, I_{5/2}(\chi_e,\tau) \right]
\ , \label{PexT}\\[0.2truecm]
\frac{{\cal E}_e(T)}{{\cal E}_0} &=&  \sqrt{2}\, \tau^{5/2}\,
\left[ I_{3/2}(\chi_e,\tau)+ \tau\, I_{5/2}(\chi_e,\tau) \right]
\ ,  \quad  {\cal E}_0= 3\rho_0\, m_ec^2\ . \label{EexT}
\eea
Now, we will need the following Sommerfeld decompositions
of the generalized Fermi-Dirac integrals
(denoting $\varphi= \tau\, \chi= \mu_e/(m_ec^2)$):
\bea
I_{1/2}(\chi,\tau) &\simeq&  \frac{1}{\sqrt{2}\, \tau^{3/2}}\,
\left\{ \frac{1}{2}\, \left[
(1+\varphi)\sqrt{\varphi(2+\varphi)}
-  {\rm ln}\left(1+\varphi + \sqrt{\varphi(2+\varphi)}\right)\, \right] + \right.
\nn\\[0.1truecm]
&& \hspace*{2.0truecm}
\left. + \frac{(1+\varphi)}{\sqrt{\varphi(2+\varphi)}}\, \frac{\pi^2 \tau^2}{6}+
\frac{(1+\varphi)}{[\varphi(2+\varphi)]^{5/2}}\, \frac{7 \pi^4 \tau^4}{120}
+ \dots \right\} \ ,   \label{SSG05res}\\[0.1truecm]
I_{3/2}(\chi,\tau) &\simeq&  \frac{1}{\sqrt{2}\, \tau^{5/2}}\,
\left\{ \frac{2\varphi^2+\varphi-3}{6}\, \sqrt{\varphi(2+\varphi)}
+ \frac{1}{2} {\rm ln}\left(1+\varphi + \sqrt{\varphi(2+\varphi)}\right)
\right. \nn\\[0.1truecm]
&& \hspace*{2.0truecm}
\left. + \frac{\sqrt{\varphi}\, (3+2\varphi)}{\sqrt{2+\varphi}}\,
\frac{\pi^2 \tau^2}{6}
- \frac{1}{\varphi^{3/2}\, (2+\varphi)^{5/2}}\,
\frac{7 \pi^4 \tau^4}{120} + \dots   \right\}   \ , \label{SSG35res}\\[0.1truecm]
I_{5/2}(\chi,\tau) &\simeq&  \frac{1}{\sqrt{2}\, \tau^{7/2}}\,
\left\{ \frac{(6\varphi^3+2\varphi^2-5\varphi+15)\sqrt{\varphi(2+\varphi)}}{24}\,
- \frac{5}{8} {\rm ln}\left(1+\varphi + \sqrt{\varphi(2+\varphi)}\right)
\right. \nn\\[0.1truecm]
&& \hspace*{0.5truecm}
\left. + \frac{\varphi^2\, (3\varphi+5)}{\sqrt{\varphi(2+\varphi)}}\,
\frac{\pi^2 \tau^2}{6}
+ \frac{\varphi^2(2\varphi^3+ 10\varphi^2+ 15\varphi+5)}{[\varphi(2+\varphi)]^{5/2}}\,
\frac{7 \pi^4 \tau^4}{120} + \dots   \right\}   \ . \label{SSG55res}
\eea
Now we again first find the chemical potential from the condition
$n_e(T)= n_e(0)$. Substituting (\ref{SSG05res},\ref{SSG35res})
into equation (\ref{rhoexT}) for $n_e(T)$  yields
(the logarithmic terms cancel each other):
\bea
&& \hspace*{-1.9truecm}
\frac{n_e(T)}{\rho_0}  \simeq   \left[ \varphi(2+\varphi) \right]^{3/2}+
\frac{(2\varphi^2+4\varphi+1)}{\sqrt{\varphi(2+\varphi)}}\, \frac{\pi^2 \tau^2}{2}+
\frac{1}{[\varphi(2+\varphi)]^{5/2}}\, \frac{7 \pi^4 \tau^4}{40}+
\dots  \ . \label{rhoexTS}
\eea
For $T=0$ it holds $\tau=0$ and
\bea
\varphi(0) \equiv \varphi_0= \frac{\mu_0}{m_ec^2}=
\tilde{\epsilon}_F= \epsilon_F-1 \ , \quad  \epsilon_F= \sqrt{1+x^2_F} \ . \nn
\eea
This relation implies:
\bea
\varphi_0(2+\varphi_0)=(\epsilon_F-1)(\epsilon_F+1)=
\epsilon_F^2-1= x^2_F \ , \nn
\eea
which reproduces the electron density at zero temperature:
from the relation (\ref{rhoexTS}) one gets
\bea
n_e(0) &=& \rho_0\, \left[ \varphi_0(2+\varphi_0) \right]^{3/2}=
\rho_0\, x^3_F \equiv \rho_F \ . \nn
\eea

Using (\ref{rhoexTS}) one gets from $n_e(0)= n_e(T)$
an implicit relation between $\varphi_0$ and $\varphi$:
\bea
&& \hspace*{-2.0truecm}
\varphi_0(2+\varphi_0) =
\varphi(2+\varphi) \,
\left[1+
\frac{(2\varphi^2+4\varphi+1)\, B}{2[\varphi(2+\varphi)]^2}
+ \frac{7\, B^2}{40[\varphi(2+\varphi)]^4} + \dots \right]^{2/3}
\ , \label{rhorho0}
\eea
where $B= \pi^2\, \tau^2$. To solve this constraint we
assume for  $\varphi= \mu_e/(m_ec^2)$
a perturbative expansion in a form:
\bea
\varphi &=& \varphi_0\, \left(1+ c_1\frac{B}{\varphi^2_0}
+ c_2\frac{B^2}{\varphi^4_0}+ \dots \right) \ . \label{varphiAns}
\eea
Notice that:
\bea
\frac{B}{\varphi^2_0}&=&
\frac{\pi^2 (k_B T)^2}{(m_ec^2)^2}\, \frac{(m_ec^2)^2}{\mu_0^2}=
\frac{\pi^2 (k_B T)^2}{\mu_0^2} \ , \nn
\eea
hence our Ansatz for $\varphi$ above is identical to the Ansatz
used for $\mu_e$ in previous sections. Substituting (\ref{varphiAns})
into (\ref{rhorho0}), making the Taylor decomposition in powers
of $B$ and requiring that the coefficients in front of $B^n, n\geq 1$ are
equal to zero,  yields the coefficients $c_i$ in terms of $\varphi_0$.
The first two are given by relatively simple equations:
\bea
c_1 &=& - \frac{2\varphi^2_0+ 4\varphi_0+1}{6(\varphi^2_0+ 3\varphi_0+2)}=
- \frac{2\epsilon^2_F-1}{6\epsilon_F(\epsilon_F+1)} \ , \label{c1gen}\\[0.2truecm]
c_2&=& - \frac{20\varphi^4_0 + 80\varphi^3_0+ 141 \varphi^2_0+ 122\varphi_0+36}
{360(\varphi^2_0+ 3\varphi_0+2)^3}
= -\frac{20\epsilon^4_F+21\epsilon^2_F-5}{360\epsilon^3_F(\epsilon_F+1)^3} \ . \label{c2gen}
\eea
With these $c_1$ and $c_2$ in (\ref{varphiAns}) the Taylor
decomposition of (\ref{rhorho0}) has a first non-zero coefficients
(apart from the constant $\varphi_0(2+\varphi_0)$ term) in front of the power
$B^3$.

Let us now derive the perturbative series for
the pressure and the kinetic energy.
Substituting the decompositions of $I_{3/2}(\chi_e,\tau)$
(see \ref{SSG35res}) and $I_{5/2}(\chi_e,\tau)$ (see \ref{SSG55res})
into (\ref{PexT})  and   (\ref{EexT}) yields:
\bea
\frac{P(T)}{\rho_0\, m_ec^2}  &=&
\frac{(2\varphi^3+6\varphi^2+\varphi-3)\sqrt{\varphi(2+\varphi)}}{8}
+  \frac{3}{8}\, {\rm ln}\left(1+\varphi + \sqrt{\varphi(2+\varphi)}\right)
\label{PexTS}\\[0.2truecm]
&&
+ \frac{(\varphi^3+3\varphi^2+2\varphi)}{\sqrt{\varphi(2+\varphi)}}\,
\frac{\pi^2 \theta^2}{2}+
\frac{(2\varphi^5+10\varphi^4+15\varphi^3+5\varphi^2-2\varphi)}{[\varphi(2+\varphi)]^{5/2}}\,
\frac{7 \pi^4 \theta^4}{120}
+ \dots \ , \nn\\[0.4truecm]
\frac{{\cal E}_e(T)}{{\cal E}_0}&=&
\frac{(6\varphi^3+10\varphi^2-\varphi+3)\sqrt{\varphi(2+\varphi)}}{24}
-  \frac{1}{8}\, {\rm ln}\left(1+\varphi + \sqrt{\varphi(2+\varphi)}\right)
\label{EexTS}\\[0.2truecm]
&&
+ \frac{(3\varphi^3+7\varphi^2+3\varphi)}{\sqrt{\varphi(2+\varphi)}}\,
\frac{\pi^2 \theta^2}{6}+
\frac{(2\varphi^5+10\varphi^4+15\varphi^3+5\varphi^2-\varphi)}{[\varphi(2+\varphi)]^{5/2}}\,
\frac{7 \pi^4 \theta^4}{120} + \dots \ . \nn
\eea
What remains to be done is to substitute $\varphi$ expressed in terms of
$\varphi_0$ and powers of $B$ (see eqs.~(\ref{varphiAns})-(\ref{c2gen}))
into equations for the pressure (\ref{PexTS}) and kinetic energy (\ref{EexTS})
and decompose into the Taylor series in powers of $B$. This is not as
simple as for the limiting non-relativistic and ultra-relativistic cases
considered above, since the coefficients depend on $\varphi_0$ and also $P(0)$
and ${\cal E}(0)$ are now not just simple powers of $\varphi_0$ and
cannot be simply factorized. Nevertheless, with the help of \emph{Mathematica}
the power decomposition can be performed. We made a decomposition up to
the order $B^2$, but the terms $\sim B^2$ are  is lengthy and clumsy.
hence, we for the sake of briefness present just leading and next-to-leading orders:
\bea
\frac{P(T)}{\rho_0\, m_ec^2} &=& \frac{P(0)}{\rho_0\, m_ec^2}+
\frac{\varphi_0(2+\varphi_0)(\varphi_0^2+2\varphi_0+2)\, B}
{6(1+\varphi_0)\sqrt{\varphi_0(2+\varphi_0)}} \nn\\
&=& \frac{P(0)}{\rho_0\, m_ec^2}+ \frac{x_F(\epsilon_F^2+1)}{6\epsilon_F}\, B \ ,
\quad B= \pi^2\, \tau^2 \ , \label{PexSomm}\\[0.2truecm]
\frac{{\cal E}_e(T)}{{\cal E}_0} &=& \frac{{\cal E}_e(0)}{{\cal E}_0}+
\frac{\varphi_0(2+\varphi_0)(1+\varphi_0)\, B}
{6\sqrt{\varphi_0(2+\varphi_0)}} = \frac{{\cal E}_e(0)}{{\cal E}_0}+ \frac{x_F\, \epsilon_F}{6}\, B \ .
\label{EexSomm}
\eea

\subsection{Sommerfeld decomposition of entropy}

Recall the equation for the dimensionless reduced entropy of
free electrons (\ref{hats}):
\bea
\hat{s}_e &\equiv& \frac{1}{k_B}\, \frac{S_e}{N_e}\equiv
\frac{1}{k_B\, T}\, \Sigma \ , \quad
\Sigma=   \frac{{\cal E}_e+ P_e}{n_e}- \mu_e  \ , \label{sereddef2}
\eea
where it is convenient to separate for a while a factor $\Sigma$.
We re-write this factor in a convenient form:
\bea
\Sigma&=&  \frac{{\cal E}_e+ P_e}{n_e}-  \mu_e =
\left[ \frac{{\cal E}_e(0)+ P_e(0)}{n_e}-  \mu_e  \right]+
\frac{\Delta {\cal E}_e+ \Delta P_e}{n_e}=
\Sigma_1+ \Sigma_2 \ , \label{sigmat}
\eea
where $\Delta {\cal E}_e={\cal E}_e(T)- {\cal E}_e(0)$
and $\Delta P_e=  P_e(T)-  P_e(0)$. From relations (\ref{PexTS},\ref{EexTS})
at $T=0$ and from $\varphi(0)\equiv \varphi_0= \tilde{\epsilon}_F$ one gets:
\bea
{\cal E}_e(0)+ P_e(0)&=& m_ec^2\, \rho_0\, x^3_F\, \tilde{\epsilon}_F=
m_ec^2\, n_e\, \tilde{\epsilon}_F \ \rightarrow \
\frac{{\cal E}_e(0)+ P_e(0)}{n_e}= m_ec^2\, \varphi_0 \ .
\nn
\eea
Making use of $ \mu_e= m_ec^2\, \varphi$ we write the first term of
$\Sigma$ in a compact form:
\bea
\Sigma_1 &=&  \frac{{\cal E}_e(0)+ P_e(0)}{n_e}-  \mu_e =
- m_e\, c^2\, (\varphi- \varphi_0)   \ . \label{sigma1}
\eea
The low temperature decomposition of this equation follows from (\ref{varphiAns}):
\bea
\varphi- \varphi_0&\simeq& c_1\, \frac{B}{\varphi_0}=
- \frac{2\epsilon^2_F-1}{6\epsilon_F\, x^2_F}\, B
\ , \nn
\eea
which implies:
\bea
\Sigma_1&\simeq &  m_e\, c^2\, \frac{2\epsilon^2_F-1}{6\epsilon_F\, x^2_F}\, B \ ,
\quad B= \pi^2\, \tau^2 \ . \label{sigma1a}
\eea
The non-relativistic ($\epsilon_F\rightarrow 1$) and ultra-relativistic
($\epsilon_F\rightarrow \tilde{\epsilon}_F \rightarrow x_F$) limits of
this equation read:
\bea
\Sigma_{1,nr}= m_e\, c^2\, \frac{1}{6\, x^2_F}\, B \ , \quad
\Sigma_{1,ur}= m_e\, c^2\, \frac{1}{3\, x_F}\, B \ .
\eea
These equations can be also obtained by calculating $\varphi- \varphi_0$
directly from (\ref{tilmunrT}) and (\ref{tilmuurT}).

The second part of $\Sigma$ is obtained from (\ref{PexSomm},\ref{EexSomm}):
\bea
\Delta {\cal E}_e+ \Delta P_e&=&
m_ec^2\, \rho_0\,  \left( \frac{x_F\epsilon_F}{2}
+\frac{x_F(\epsilon^2_F+1)}{6\epsilon_F} \right)\,
B= m_ec^2\, \rho_0\,
\frac{x_F\, (4\epsilon^2_F+1)}{6\epsilon_F}\, B \ , \nn
\eea
which implies:
\bea
\Sigma_2&=&  \frac{\Delta {\cal E}_e+ \Delta P_e}{n_e}=
m_ec^2\,  \frac{4\epsilon^2_F+1}{6\epsilon_F\,x^2_F}\, B \ . \label{sigma2a}
\eea
In kinematic limits the factor $\Sigma_2$ reduces to
\bea
\Sigma_{2,nr}= m_e\, c^2\, \frac{5}{6\, x^2_F}\, B \ , \quad
\Sigma_{2,ur}= m_e\, c^2\, \frac{2}{3\, x_F}\, B \ .
\eea

Adding up the results for $\Sigma_1$ and $\Sigma_2$ yields the total $\Sigma$:
\bea
\Sigma&=& k_BT\, \hat{s}_e \equiv \frac{T S_e}{N_e}=
m_ec^2\,  \frac{\epsilon_F}{x^2_F}\, B \ , \quad B= \pi^2\, \tau^2 \ ,
\label{sigma_somm}\\[0.1truecm]
\Sigma_{nr}&=& m_ec^2\,  \frac{1}{x^2_F}\, B \ , \qquad
\Sigma_{ur}= m_ec^2\,  \frac{1}{x_F}\, B \ . \nn
\eea
Equations on the 2nd line can be, of course, obtained directly from
the non-relativistic or ultra-relativistic results for the momentum
and energy densities.

\end{document}